\documentclass[lettersize,journal]{IEEEtran}
\usepackage{amsmath,amsfonts}
\usepackage{array}
\usepackage{textcomp}
\usepackage{stfloats}
\usepackage{url}
\usepackage{subfigure}
\usepackage{verbatim}
\usepackage{graphicx}
\usepackage{cite,color}
\usepackage{stmaryrd}
\usepackage{algpseudocode}
\usepackage{caption}
\usepackage{ragged2e}
\usepackage{booktabs}
\usepackage[ruled,linesnumbered]{algorithm2e}
\usepackage{amsmath,amssymb,amsfonts}

\allowdisplaybreaks

\usepackage{xcolor}
\definecolor{MXY}{RGB}{252,8,235}

\usepackage[font={small,it}]{caption}
\begin{document}

\title{Multi-Hop RIS ISAC for Target Positioning:\\ A Tensor Decomposition-based Approach}

\author{Yirui Luo,~\IEEEmembership{Student Member,~IEEE,} Xiaoyan Ma,~\IEEEmembership{Member,~IEEE},
Yong Liang Guan,~\IEEEmembership{Senior Member,~IEEE}, \\
Christopher G. Brinton,~\IEEEmembership{Senior Member,~IEEE}, and Chau Yuen,~\IEEEmembership{Fellow,~IEEE}
	\thanks{ Yirui Luo, Yong Liang Guan and Chau Yuen are from School of Electrical and Electronic Engineering, Nanyang Technological University, Email: yirui001@e.ntu.edu.sg, \{eylguan, chau.yuen\}@ntu.edu.sg. Xiaoyan Ma and Christopher G. Brinton are from School of Electrical and Computer Engineering, Purdue University, Email: \{ma946, cgb\}@purdue.edu.
    Yirui Luo and Xiaoyan Ma contributed equally to this work.
	}
}

\maketitle

\begin{abstract}
Reconfigurable intelligent surface (RIS) has demonstrated remarkable potential to enhance the performance of integrated sensing and communication (ISAC), particularly when the line-of-sight (LoS) paths are obstructed. By controlling the reconfigurable elements on the surface, RIS can establish virtual LoS paths and provide considerable passive beamforming gains, thereby significantly improving the received signal quality. In this paper, we design a novel {multi-hop RIS} ISAC system for target positioning, where multiple RISs are deployed to assist the communication from a transmitter to associated users while simultaneously enhancing receiver sensing performance in target positioning. Specifically, we formulate an optimization problem to minimize the root mean square error (RMSE) of the target detection while guaranteeing the communication requirements of the users. To solve this problem, we first unfold the cascaded sensing channel through parallel factor decomposition, and develop a low-rank CANDECOMP/PARAFAC decomposition (CPD)-based scheme to extract the location parameters (i.e., angle of arrival, angle of departure and delay) of the sensing targets. Then, we develop a scheme for jointly selecting the transmit beamforming and RIS phase shift configurations to maximize the sensing energy at the receiver, which in turn leads to improved accuracy in target positioning. We also provide a uniqueness analysis, complexity analysis, and Cramér-Rao lower bound (CRLB) of the parameters estimated by our methodology. Simulation results validate the improvement in target positioning obtained by our design relative to baselines. 

\end{abstract}


\section{Introduction}
In the sixth-generation (6G) wireless networks, various emerging services, such as vehicle-to-everything (V2X), virtual reality (VR), and precision security monitoring call for higher-speed communications together with accurate and reliable sensing capabilities \cite{Nguyen2022,Z_Zhang2019,Q_Qi2022,brinton2025key}. In response to such demand, both wireless communication and radar sensing systems have been continuously expanding their frequency bands, thereby increasing the shortage of available spectrum resources \cite{10319318}. To cope with this problem, integrated sensing and communication (ISAC) technology is regarded as a promising solution. By integrating wireless communication and radar sensing into the same system and allowing them to co-exist at the same frequencies, ISAC can significantly enhance spectral and energy efficiencies \cite{ruoyu_ISAC}. Additionally, its sensing capabilities can potentially enable security features associated with monitoring and localization~\cite{zhu2024enabling,cao2024sensing}, which is a major motivating scenario for this work. 

Among the various strategies for ISAC, one compelling approach is to make use of the existing communication hardware and signals through advanced techniques in sensing signal processing \cite{9540344}. For example, the authors in \cite{80211_ad_8114253} performed range and velocity estimation by leveraging standard wireless local area network (LAN) receivers and classical pulse-Doppler algorithms. Alternatively, the orthogonal frequency division multiplexing (OFDM) modulation technique has also been employed in ISAC scenarios. OFDM can enhance the robustness of target detection and the accuracy of parameter estimation by leveraging frequency diversity, thereby providing improved sensing capability \cite{Novel_10634583}. Many studies have utilized OFDM communication signals in this way, e.g., work \cite{OFDM_5776640} proposed a target parameter estimation method based on symbol division using OFDM signal. To improve the estimation accuracy, super-resolution algorithms were studied in \cite{superresolution_7833233,yirui_10200722,yongjun_9166743}. Moreover, by using DFT-spread OFDM, a deep-learning-based algorithm for terahertz (THz) systems was proposed in \cite{THz_9967989}. The large bandwidth available in the millimeter wave (mmWave) or THz frequency bands can assist ISAC systems in achieving high data rates and high sensing resolution. However, {severe propagation loss in these bands and obstruction from environmental objects often hinder line-of-sight (LoS) paths,} thus degrading the performance of ISAC and target positioning capabilities.

A common goal of ISAC, particularly in these target positioning applications, is to extract location information of sensing targets \cite{su2023sensing,brinton2025key}. The receiver typically needs to estimate target-related parameters from the received signals. To achieve this, the authors in \cite{liu_fan_8999605} proposed the MUltiple SIgnal Classification (MUSIC) method for angle estimation and a matched-filtering (MF) method for delay and Doppler shift estimation. By utilizing sparsity in the angle domain, the authors in \cite{Lee_7458188,9573305} proposed compressed sensing (CS)-based methods for channel estimation, which also contains angle of arrival (AoA), angle of departure (AoD), and delay information. Recently, tensor decomposition has received wide attention for high-dimensional signal processing, as it enables exploiting multidimensional characteristics for parameter matching and performance enhancement. In particular, by utilizing the sparsity of the channel, the authors in \cite{low_rank} and \cite{MYIOTJ10963873} proposed the CANDECOMP/PARAFAC decomposition (CPD)-based method for channel and target parameter estimation. However, {these parameter estimation methods are based on the assumption that the targets are in the LoS of the transmitter.} They cannot be applied in non-LoS (NLoS) settings directly, as is often the case with indoor (e.g., home) monitoring applications. 

Fortunately, the emergence of reconfigurable intelligent surface (RIS) technology offers a potential solution to the LoS challenges encountered in ISAC parameter estimation \cite{GAN_10543050,li10438390}. By adjusting the phase shift and amplitude of low-power, cost-effective controllable reflective elements, RIS can flexibly reconfigure the wireless propagation environment to establish virtual LoS paths \cite{xiaoyan_9829192} and improve the overall performance of the system \cite{Rihan_10194901,xiaoyan_9320618,chenjie10053657}. In \cite{Ris1}, the authors investigate the integration of
 ISAC and RIS for providing wide coverage, ultra
reliable communication, and high-accuracy sensing functions. Their formulated optimization problem aims to maximize the sum-rate of the communication users while guaranteeing a worst case signal-to-noise (SNR) constraint of sensing requirements. In addition to SNR sensing requirements, the authors in \cite{RIS2} focused on the beampattern similarity aspect of sensing. In their work, the active beamforming of the BS and the passive beamforming of the RIS are jointly optimized to maximize the achievable sum-rate of the communication users while satisfying a constraint on beampattern similarity for radar sensing. In \cite{RIS3}, the authors investigate the interference problem and propose a dual-functional waveform design
 scheme for RIS-aided ISAC to minimize the  weighted sum of  multi-user interference energy and waveform discrepancy.

Taken together, these results imply that integrating RIS into ISAC can provide significant benefits for both communication and sensing. However, while the existing RIS-related ISAC works have considered sensing-related metrics like SNR and beampattern similarity, these methods do not provide specific locations of sensing targets, which is critical in security monitoring applications.
Motivated by this, in this work, we develop a novel target positioning methodology for {multi-hop RIS-assisted ISAC}, specifically considering the problem of home security monitoring. In our system, multiple RISs are deployed to assist the communication from the transmitter to associated communication users while simultaneously enhancing the receiver’s sensing performance in localizing a target of interest. Leveraging sparsity in the sensing channels, we develop a tensor decomposition-based target parameter estimation method that employs the CANDECOMP/PARAFAC (CP) model \cite{low_rank,MYIOTJ10963873}. We also propose an optimization methodology for minimizing the localization error by adjusting the beamforming vectors of the transmitter and the phase shift designs of the RISs. 

\subsection*{Outline and Summary of Contributions}
The main contributions of the paper are summarized as follows:

\begin{itemize}
\item We propose a novel {multi-hop RIS} ISAC system for target positioning in home security monitoring scenario, where the communication signals are reused by the sensing receiver to localize outdoor targets of interest. In our setting, we assume pre-deployment of multiple RISs in the home, enabling the creation of virtual LoS links for communication purposes while improving the accuracy of localization results.

\item We design a CP decomposition (CPD)-based sensing scheme to extract location-related parameters of the targets from signals received at the sensing receiver. By exploiting the inherent multidimensional structure of the signal and the sparsity of the channel, we characterize the received sensing signal as a third-order tensor that fits the CP model with embedded AoD, AoA and delay parameters. Through CP decomposition of the tensor, we are able to estimate the multi-target parameters from its factor matrices. Additionally, we analyze the uniqueness of the CP decomposition to provide accuracy guarantees on the estimated parameters.

\item We formulate a joint transmit beamformer and RIS phase shift optimization problem to minimize the target localization error while ensuring user communication requirements. To overcome the complexity of the optimization, we decouple the original problem into two sub-problems, one for the beamforming design and the other for the phases at each RIS. We further demonstrate how each sub-problem can be transformed into a convex problem, thereby readily solvable through existing optimization algorithms. Additionally, we provide a formal complexity analysis of our entire methodology.
{\item Finally, we conduct a numerical evaluation of our proposed beamforming with the proposed CPD-based sensing scheme. We show that the proposed sensing scheme achieves fast convergence and its performance approaches the corresponding Cramér-Rao lower bound (CRLB) on estimation error. We also provide comparisons of the localization error performance and the complexity between our proposed scheme and state-of-the-art techniques, showing that our methodology obtains substantial improvements. {What's more, we further consider imperfect cancellation of reflected signals from communication users by introducing a residual cancellation factor, and demonstrate the superiority of our proposed beamforming scheme under different residual cancellation levels and communication SNR thresholds.}}
\end{itemize}

The remainder of the paper is organized as follows. Section II presents the signal model and the channel model for the {multi-hop RIS} ISAC home security monitoring system. The proposed tensor-based target parameter estimation problem is discussed in Section III. Section IV introduces the joint beamforming and phase shift design methodology for optimizing the sensing performance.
Section V provides the complexity analysis and CRLB of our proposed scheme. 
Section VI presents the numerical results, followed by a conclusion in Section VII.

\textbf{Notation:} Lowercase letter $x$, lowercase bold letter ${\boldsymbol{x}}$, and uppercase bold letter ${\mathbf{X}}$ notate variables, vectors, and matrices, respectively. The calligraphy letter $\boldsymbol{\mathcal{X}}$ represents a tensor. ${\left(  \cdot  \right)^{ - 1}}$, ${\left(  \cdot  \right)^{\dag}}$, ${\left(  \cdot  \right)^{ *}}$, ${\left(\cdot\right)^{T}}$, ${\left(\cdot\right)^{H}}$ and $|\cdot|$ denote the inverse, pseudo-inverse, conjugate, transpose, conjugate transpose and modulus operators, respectively. $ \circ $, $ \odot $, and $* $ stand for the outer product, Khatri-Rao product, and Hadamard product, respectively. ${\left\|  \cdot  \right\|}$ and ${\left\|  \cdot  \right\|_F}$ stand for Euclidean norm and Frobenius norm, respectively. A complex random variable $x$ following a complex Gaussian distribution with mean $\mu$ and variance $\sigma^2$ is denoted $x\backsim \mathcal{CN}(\mu,\sigma^2)$. $\mathrm{Tr}\left(\mathbf{X}\right)$ represents the trace of matrix $\mathbf{X}$, $\mathrm{diag}\left(\mathbf{X}\right)$ represents the vector which contains the main diagonal elements of $\mathbf{X}$, and $\mathrm{ones}\left(N,1\right)$ represents a $N \times 1$ all-ones vector.

\section{System Model}
We consider a {multi-hop} RIS-aided ISAC system for a home security monitoring scenario. 
Specifically, a transmitter (Tx) installed in the main room provides communication services 
to both indoor and outdoor users with the assistance of multiple RISs. The indoor RISs are 
deployed to enhance the coverage of indoor NLoS areas, while the outdoor 
RISs extend the communication coverage to outdoor NLoS areas. Meanwhile, for security 
monitoring, the system aims to sense and localize passive targets appearing in front of the 
door. An outdoor sensing receiver (Rx), connected to the Tx through a wired feedback link, 
reuses the communication signals for target localization. Since the transmitted signals are 
known at the sensing Rx, the target-related parameters can be estimated from the received 
echo signals. Our objective is to minimize the root mean square error (RMSE) of multi-target 
localization while guaranteeing the communication requirements of all users. {To clearly illustrate the considered system model, Fig.~1 shows a representative two-RIS 
scenario. In this example, RIS~1 assists the Tx in serving indoor NLoS communication users 
and also forwards the signal toward RIS~2. Then, RIS~2 further supports outdoor NLoS 
communication users and enables the sensing of passive targets located near the front door. Although Fig.~1 illustrates a two-RIS home security monitoring example, the proposed framework and problem formulation can be applied to general multi-hop RIS-aided ISAC systems with an arbitrary number of RISs as shown in the following problem formulations.}


\subsection{Communication Model}
\begin{figure}[t]
\centering
\includegraphics[width=\linewidth]{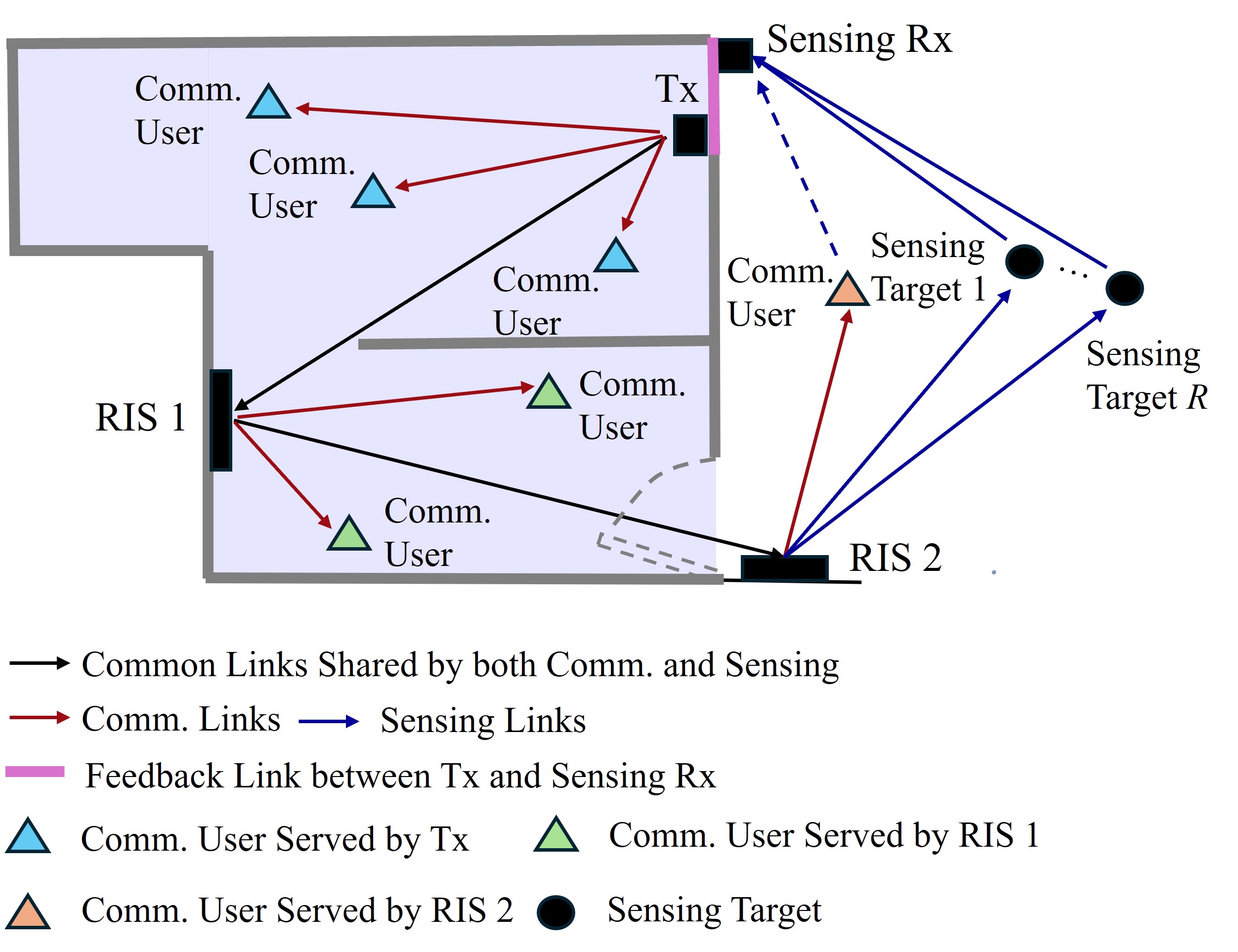}
\caption{Double-hop RIS ISAC system for home security monitoring scenario. The objective is to sense and localize the targets that appear at {the front door} using the communication signals generated by Tx in the home.}
\label{ISAC}
\end{figure}

{In this paper, we consider mmWave communications, which have been widely adopted in both indoor and outdoor scenarios \cite{FarF0, FarF1}. According to the results in \cite{FarF0}, most indoor deployment scenarios operate in the far-field region. Therefore, this paper focuses on far-field mmWave communication systems.} {Also, we adopt a LoS-dominant mmWave channel model to emphasize the impact of the RIS configuration in this work. This choice is motivated by empirical measurements in \cite{LoSR}, which demonstrate that at frequencies above the mmWave bands, the received power of the LoS path significantly exceeds that of the NLoS paths.}

{Specifically, we consider a total of $K$ communication users in the generalized multi-RIS ISAC system. 
Let $\mathcal{I}=\{1,2,\ldots,I\}$ denote the set of RISs, where RIS $i$ is equipped with 
$M_i$ reflecting elements. The users can be served either directly by the Tx or through 
multi-hop RIS-assisted links. We denote by $\mathbb{K}_0$ the set of LoS users directly 
served by the Tx, and by $\mathbb{K}_{\rm RIS}$ the set of NLoS users served through one 
or multiple RISs, with $\mathbb{K}_0\cup\mathbb{K}_{\rm RIS}=\mathbb{K}$ and 
$\mathbb{K}_0\cap\mathbb{K}_{\rm RIS}=\emptyset$. For each RIS-assisted user 
$k\in\mathbb{K}_{\rm RIS}$, we define its ordered RIS path as
$\mathcal{P}_k=(i_{k,1},i_{k,2},\ldots,i_{k,L_k})$, where $i_{k,\ell}\in\mathcal{I}$ 
denotes the $\ell$-th RIS along the propagation path from the Tx to user $k$, and $L_k$ 
is the number of RISs involved in serving user $k$. For direct users, we set 
$\mathcal{P}_k=\emptyset$.
We assume an orthogonal frequency division multiple access (OFDMA) system with an 
available bandwidth of $B=N\Delta f$, where $N$ is the number of subcarriers and 
$\Delta f$ is the subcarrier spacing. The bandwidth is equally allocated across the 
$K$ communication users, and inter-user interference is eliminated by the use of OFDMA. 
And we also assume that the Tx is equipped with $M_t$ transmit antennas.}

{The communication channel between the Tx and user $k$ depends on the corresponding 
serving path \cite{xiaoyan_9829192}. Specifically, the equivalent channel for user $k$ on subcarrier $j$ is 
given by
\begin{equation} \label{channel_general}
\boldsymbol{h}_{k,j}=
\begin{cases}
\boldsymbol{g}_{Tk,j}^{H}, 
& k\in\mathbb{K}_0,\\[1mm]
\boldsymbol{g}_{i_{k,L_k}k,j}^{H}\boldsymbol{C}_{k,j},
& k\in\mathbb{K}_{\rm RIS},
\end{cases}
\end{equation}
where $\boldsymbol{C}_{k,j}$ denotes the cascaded channel from the Tx to the last RIS 
on the serving path of user $k$, which is defined as
\begin{align}
\boldsymbol{C}_{k,j}
&=
\boldsymbol{\Phi}_{i_{k,L_k}}
\boldsymbol{G}_{i_{k,L_k-1}i_{k,L_k},j}
\boldsymbol{\Phi}_{i_{k,L_k-1}}
\cdots \nonumber\\
&\quad \times
\boldsymbol{G}_{i_{k,1}i_{k,2},j}
\boldsymbol{\Phi}_{i_{k,1}}
\boldsymbol{G}_{Ti_{k,1},j}.
\label{cascaded_channel}
\end{align}
When $L_k=1$, the serving path contains only one RIS, and thus no inter-RIS channel 
is involved. In this case, the cascaded channel reduces to
\begin{equation}
\boldsymbol{C}_{k,j}
=
\mathbf{\Phi}_{i_{k,1}}
\mathbf{G}_{Ti_{k,1},j},
\end{equation}
and the equivalent channel becomes
\begin{equation}
\boldsymbol{h}_{k,j}
=
\boldsymbol{g}_{i_{k,1}k,j}^{H}
\mathbf{\Phi}_{i_{k,1}}
\mathbf{G}_{Ti_{k,1},j}.
\end{equation}
In \eqref{channel_general}, $\boldsymbol{h}_{k,j}\in\mathbb{C}^{1\times M_t}$ denotes 
the equivalent channel from the Tx to user $k$ on subcarrier $j$, where 
$j=1,2,\ldots,J_k$ and $J_k$ is the number of subcarriers allocated to user $k$. 
We assume that the total number of subcarriers $N$ is equally allocated to all users, 
i.e., $J_1=J_2=\cdots=J_K=J=N/K$. 
For direct users, $\boldsymbol{g}_{Tk,j}\in\mathbb{C}^{M_t\times 1}$ denotes the channel 
from the Tx to user $k$ on subcarrier $j$. For RIS-assisted users, the vector 
$\boldsymbol{g}_{i_{k,L_k}k,j}\in\mathbb{C}^{M_{i_{k,L_k}}\times 1}$ denotes the channel 
from the last RIS on the path, i.e., RIS $i_{k,L_k}$, to user $k$.
Moreover, $\boldsymbol{C}_{k,j}\in\mathbb{C}^{M_{i_{k,L_k}}\times M_t}$ denotes the 
cascaded channel from the Tx to the last RIS on the serving path of user $k$. 
Specifically, $\mathbf{G}_{Ti,j}\in\mathbb{C}^{M_i\times M_t}$ denotes the channel from 
the Tx to RIS $i$, and 
$\mathbf{G}_{i\ell,j}\in\mathbb{C}^{M_{\ell}\times M_i}$ denotes the channel from RIS 
$i$ to RIS $\ell$ on subcarrier $j$. The phase-shift matrix of RIS $i$ is denoted by 
$\mathbf{\Phi}_i\in\mathbb{C}^{M_i\times M_i}$ and is defined as
$\mathbf{\Phi}_i=\mathrm{diag}(\boldsymbol{\phi}_i)$, where 
$\boldsymbol{\phi}_i=[\phi_{i,1},\phi_{i,2},\ldots,\phi_{i,M_i}]^T$ with 
$\phi_{i,m}=e^{-\mathrm{j}\theta_{i,m}}$ and $\theta_{i,m}\in[0,2\pi)$.}

Since the Tx has multiple antennas, a beamforming vector 
$\boldsymbol{w}_{k,j}\in\mathbb{C}^{M_t\times 1}$ is applied to the information symbol 
$x_{k,j}$, where $\mathbb{E}[|x_{k,j}|^2]=1$. The transmitted signal intended for user 
$k$ on subcarrier $j$ is given by 
$\boldsymbol{s}_{k,j}=\boldsymbol{w}_{k,j}x_{k,j}$. Therefore, the received signal at 
user $k$ on subcarrier $j$ can be expressed as
\begin{equation}
    y_{k,j}= \boldsymbol{h}_{k,j}\boldsymbol{w}_{k,j}x_{k,j}+n_{k,j},
\end{equation}
where $n_{k,j}\sim\mathcal{CN}(0,\sigma_c^2)$ denotes the complex additive white Gaussian 
noise at user $k$ on subcarrier $j$. Accordingly, the SNR of user $k$ on subcarrier $j$ is
\begin{equation}
    \gamma_{k,j}= \frac{|\boldsymbol{h}_{k,j}\boldsymbol{w}_{k,j}|^2}{\sigma_c^2}.
\end{equation}
The communication requirement is to guarantee that the SNR of each user on each allocated 
subcarrier is no smaller than a predefined threshold, i.e., 
$\gamma_{k,j}\geq\gamma_{\rm thr}$.

\subsection{Sensing Model} \label{Sen_Model}
In ISAC, we aim to reuse the communication signals to sense the targets that appear at the sensing area. To achieve this goal, we need two stages for target sensing:
\begin{itemize}
    \item \textbf{Stage 1: Wide beam scanning.} Due to the uncertainty in the presence and locations of potential sensing targets, the sensing area is initially scanned using wide beams to ensure rapid and comprehensive coverage. In this phase, we assume that a total of $Q$ phase configurations are generated to scan the entire area. Since wide beams are used at this stage, the received sensing signals are relatively weak. As a result, only  the number 
    and coarse location information of the targets is obtained. 

    \item \textbf{Stage 2: Narrow beam design.} {With preliminary target number  and coarse location information obtained in stage $1$, we design narrow beams to illuminate these targets, facilitating strong sensing signals at the sensing Rx to extract precise locations of targets.} The design goal of this phase is to minimize the estimated localization error at the sensing Rx while guaranteeing the SNR requirement of all communication users.
\end{itemize}

We assume that there are a total of $R$ targets in the sensing area, indexed by $r$. Let ${\varphi _r}$, ${\vartheta _{r}}$ and $\tau_r$ denote the AoA, AoD and total delay from \textcolor{black}{RIS $I$} to the sensing Rx for the $r$-th target. The total phase configuration matrix ${\mathbf{\Psi }} \in \mathbb{C}^{\textcolor{black}{M_I}\times Q}$ of \textcolor{black}{RIS $I$} generated during stage 1 is specified as
\begin{equation}
   \textcolor{black}{ {\boldsymbol{\Psi }} = \left[ {\boldsymbol{\phi }_{I,1},\boldsymbol{\phi }_{I,2},...,\boldsymbol{\phi }_{I,Q}} \right],}
\end{equation}
with \textcolor{black}{$\boldsymbol{\phi }_{I,q} \in \mathbb{C}^{M_I \times 1}$} representing the $q$-th phase configuration vector of \textcolor{black}{RIS $I$} for $q=1,2,...,Q$. 

\textcolor{black}{Then, the whole sensing channel $\mathbf{H}_{q,n}\in \mathbb{C}^{M_r\times M_t}$ from Tx to the sensing Rx  for the $n$-th subcarrier using the $q$-th phase configuration of RIS $I$ can be expressed as 
\begin{equation} 
\begin{split}
{{\mathbf{H}}_{q,n}} =&\sum\nolimits_{r = 1}^R {\rho _r}{{\mathbf{a}}_R}\left( {{\varphi _r}} \right){\mathbf{a}}_{I,T}^H\left( {{\vartheta _{r}}} \right){e^{ - j2\pi {f_n}{\tau _r}}}{{\mathbf{\Phi }}_{I,q}}\mathbf{G}_{TI,n}\\
&+\sum\nolimits_{k = 1}^{K_I} {\rho _{k}}{{\mathbf{a}}_R}\left( {{\varphi _{k}}} \right){\mathbf{a}}_{I,T}^H\left( {{\vartheta _{k}}} \right){e^{ - j2\pi {f_n}{\tau_{k}}}}{{\mathbf{\Phi }}_{I,q}}\mathbf{G}_{TI,n}\\
=&\mathbf{H}_{\text{SEN},q,n}+\mathbf{H}_{\text{COM},q,n},  
\end{split}\label{raw_channel}
\end{equation}
where $\mathbf{H}_{\text{SEN},q,n}\in \mathbb{C}^{M_r\times M_t}$ contains the information of unknown outdoor targets from Tx to the sensing Rx for the $n$-th subcarrier using the $q$-th phase configuration of RIS $I$ and ${{\mathbf{H}}_{\text{COM},q,n}}\in \mathbb{C}^{M_r\times M_t}$ contains the corresponding information of known active outdoor communication users with the expression of
\begin{equation} 
\begin{split}
{{\mathbf{H}}_{\text{SEN},q,n}} = &\sum\nolimits_{r = 1}^R {\rho _r}{{\mathbf{a}}_R}\left( {{\varphi _r}} \right){\mathbf{a}}_{I,T}^H\left( {{\vartheta _{r}}} \right){e^{ - j2\pi {f_n}{\tau _r}}}{{\mathbf{\Phi }}_{I,q}}\mathbf{G}_{TI,n}, 
\end{split}\label{sensing_channel}
\end{equation}
and
\begin{equation} 
\begin{split}
{{\mathbf{H}}_{\text{COM},q,n}} = &\sum\nolimits_{k = 1}^{K_I} {\rho _{k}}{{\mathbf{a}}_R}\left( {{\varphi _{k}}} \right){\mathbf{a}}_{I,T}^H\left( {{\vartheta _{k}}} \right){e^{ - j2\pi {f_n}{\tau_{k}}}}{{\mathbf{\Phi }}_{I,q}}\mathbf{G}_{TI,n}, 
\end{split}\label{comm_channel}
\end{equation}
respectively. 
$\rho_r$ and $\rho_k$ denote the complex channel gain for the $r$-th unknown target and $k$-th known active outdoor communication user from RIS $I$ to the Rx, respectively. ${\varphi _k}$, ${\vartheta _{k}}$ and $\tau_k$ denote the AoA, AoD, and total delay from RIS $I$ to the sensing Rx for the $k$-th known active outdoor communication user, respectively. $K_I$ is the number of active outdoor communication users, $f_n=n\Delta f$ is the central frequency of the $n$-th subcarrier, {${{\mathbf{\Phi }}_{I,q}}={\text{diag}}(\boldsymbol{\phi }_{I,q})$}, $\mathbf{G}_{TI,n}\in \mathbb{C}^{M_I\times M_t}$ is the cascaded channel from the Tx to RIS $I$ on the $n$-th subcarrier. ${{\mathbf{a}}_R}\left( {{\varphi }} \right)\in \mathbb{C}^{M_r\times 1}$ and ${\mathbf{a}}_{I,T}\left( {\vartheta } \right)\in \mathbb{C}^{M_I\times 1}$ are the linear array responses of sensing Rx and RIS $I$, which can be expressed as
\begin{equation}
    {\mathbf{a}_R}\left( \varphi \right) = {\left[1,{{e^{ - j\pi \sin \varphi }}},...,{{e^{ - j\pi\left( {M_r - 1} \right)\sin \varphi  }}} \right]^T},
\end{equation}
\begin{equation}
    {\mathbf{a}}_{I,T}\left( {\vartheta } \right)= {\left[1,{{e^{ - j\pi \sin {\vartheta } }}},...,{{e^{ - j\pi\left( {M_I - 1} \right)\sin \vartheta  }}} \right]^T},
\end{equation}
respectively, and {$M_r$} is the number of receiving antennas at the sensing Rx. 
}

\textcolor{black}{We further assume that during the $q$-th phase configuration of RIS $I$, $T$ symbols are transmitted.
Thus, the frequency domain received sensing signal ${{\mathbf{Y}}_{q,n}}\in \mathbb{C}^{M_r\times T}$ on the $n$-th subcarrier using the $q$-th phase configuration of RIS $I$  has an expression of
\begin{equation}
    {{\mathbf{Y}}_{q,n}} = {{\mathbf{H}}_{\text{SEN},q,n}}{{\mathbf{S}}_{q,n}}+{{\mathbf{H}}_{\text{COM},q,n}}{{\mathbf{S}}_{q,n}} + {{\mathbf{N}}_{q,n}},    
\end{equation}
where ${{\mathbf{N}}_{q,n}} \backsim \mathcal{CN}(\mathbf{0},\sigma_s^2\boldsymbol{I})$ is the noise term at the sensing Rx, which follows the complex Gaussian distribution with zero mean and variance $\sigma_s^2$.
${\mathbf{S}}_{q,n}\in \mathbb{C}^{M_t\times T}$ is the transmitted signal of the $n$-th subcarrier using the $q$-th phase configuration of RIS $I$, with $n=1,...,N$.}

\textcolor{black}{Since the outdoor communication users are known active downlink users served by the Tx and RIS-assisted network, the user-related channel state information (CSI) are available or can be estimated during the channel estimation stage. Therefore, the reflected component from these known communication users can be reconstructed at the sensing Rx as
\begin{equation}
    \mathbf{\hat Y}_{\text{COM},q,n}=\mathbf{\hat H}_{\text{COM},q,n}{{\mathbf{S}}_{q,n}}.
\end{equation}
Then, the sensing Rx subtracts the received signals reflected by the outdoor communication users from the original received sensing signals, yielding
\begin{equation} 
\begin{split} 
    \mathbf{Y}_{\text{SEN},q,n}=&{{\mathbf{Y}}_{q,n}}-\mathbf{\hat Y}_{\text{COM},q,n}\\
    =&\mathbf{H}_{\text{SEN},q,n}{{\mathbf{S}}_{q,n}}+\left(\mathbf{H}_{\text{COM},q,n}-\mathbf{\hat H}_{\text{COM},q,n}\right){{\mathbf{S}}_{q,n}}\\
    &+{{\mathbf{N}}_{q,n}}.\label{have_re_in}
\end{split}
\end{equation}
Here we assume the channel estimation for the active communication users is perfect \cite{xiaoyan_9829192,11506510} and the reconstruction for the reflected component from these known communication users is perfect. Then Eq.~\eqref{have_re_in} can be reduced to as
\begin{equation}
\mathbf{Y}_{\text{SEN},q,n}=\mathbf{H}_{\text{SEN},q,n}{{\mathbf{S}}_{q,n}}+{{\mathbf{N}}_{q,n}}.
\end{equation}}

Until now, we have formulated the received signals which contains the information about the targets at the sensing Rx. In the next section, we will explain how to obtain the location of the sensing targets from the received signal ${{\mathbf{Y}}_{\text{SEN},q,n}}$.

\section{Location Parameters Extraction through CANDECOMP/PARAFAC Decomposition  } \label{Sec_Sensing}
In this section, we provide detailed explanations about how our proposed method performs location estimation for multiple outdoor targets. Specifically, we construct the received signals into a tensor model and obtain the corresponding parameters through CP decomposition. Furthermore, we analyze the uniqueness conditions of the CP decomposition to ensure that the resulting factor matrices reliably capture the underlying target-related information.
\subsection{CP Decomposition for Factor Matrices}
Since the Tx and the sensing Rx are assumed to be connected via cables, this configuration allows the sensing Rx to have access to the original transmitted signals. By comparing the received signals with the known transmitted signals, the sensing Rx can analyze the differences caused by propagation effects. These differences are then used to extract location information about the sensing targets. After removing the transmitted information, the sensing channel estimated at the receiver can be expressed as 
\begin{equation}
\begin{split}
  {\hat{\mathbf{H}}_{\text{SEN},q,n}} &= {{\mathbf{Y}}_{\text{SEN},q,n}}{\mathbf{S}}_{q,n}^H  \\
  &=\mathbf{G}_{IR,n}{{\mathbf{\Phi }}_{I,q}}{{\mathbf{G}}_{TI,n}}+ {{{\mathbf{\tilde N}}}_{q,n}}, 
\end{split}\label{estimated_whole_channel}
\end{equation}
where $\mathbf{G}_{IR,n}=\sum\nolimits_{r = 1}^R{\rho _r}{{\mathbf{a}}_R}\left( {{\varphi _r}} \right){\mathbf{a}}_{I,T}^H\left( {\vartheta _{r}} \right){e^{ - j2\pi {f_n}{\tau _r}}}$ for $n=1,...,N$, and ${{{\mathbf{\tilde N}}}_{p,n}}={{{\mathbf{N}}}_{q,n}}{\mathbf{S}}_{q,n}^H$ is the related noise. According to Eq. \eqref{estimated_whole_channel}, the multi-dimensional channel for the $n$-th subcarrier can also be reformulated into a tensor form $\boldsymbol{\mathcal{H}}_k\in \mathbb{C}^{M_r\times M_t\times Q}$ with the expression of
\begin{equation}
    {\boldsymbol{\mathcal{H}}_n} = \left[\kern-0.15em\left[ {{{\mathbf{G}}_{IR,n}},{{\mathbf{G}}_{TI,n}},{\mathbf{\Psi }}} 
 \right]\kern-0.15em\right],
\end{equation}
since the channel from Tx to RIS $I$, i.e., $\mathbf{G}_{TI,n}$, and the beamforming matrix of RIS $I$, i.e., $\boldsymbol{\Psi }$, are already known by the transmitter and receiver, the sensing channel matrix ${\mathbf{G}}_{IR,n}$ of the $n$-th subcarrier can be obtained by 
\begin{equation}
\begin{split}
    {{\mathbf{\hat G}}_{IR,n}} &= {\left( {{{\left( {{\mathbf{\Psi }} \odot {{\mathbf{G}}_{TI,n}}} \right)}^\dag }{\mathbf{H}}_{(1,n)}^T} \right)^T}\\
    &={\mathbf{H}}_{(1,n)}\left({{\mathbf{\Psi }} \odot {{\mathbf{G}}_{TI,n}}}\right)\left({{\mathbf{\Psi }}^T{\mathbf{\Psi }} * {{\mathbf{G}}_{TI,n}}^T{{\mathbf{G}}_{TI,n}}}\right)^{-1},\label{G2k}
\end{split}
\end{equation}
where $\mathbf{H}_{(1,n)}$ is the mode-1 unfolding of tensor $\boldsymbol{\mathcal{H}}_n$. We formulate the third-order tensor $\boldsymbol{\mathcal{G}}_{IR} \in \mathbb{C}^{M_r\times M_I \times N}$, whose third dimension contains all $N$ matrices ${{\mathbf{\hat G}}_{IR,n}}$ in Eq. \eqref{G2k} for $n=1,...,N$ as
\begin{equation}
\begin{split}
    {\boldsymbol{\mathcal{G}}_{IR}} &= \sum\limits_{r = 1}^R  {{\mathbf{a}}_R}\left( {{\varphi _r}} \right) \circ {{\mathbf{a}}_{I,T}}\left( \vartheta _{r} \right) \circ \left({{\rho _r}}{\mathbf{a}_\tau}\left( {{\tau _r}} \right)\right)\\
    &= \left[\kern-0.15em\left[ {{\mathbf{A}_R},{\mathbf{A}_T},{\mathbf{A}_\tau}} \right]\kern-0.15em\right],
\end{split}\label{CPD}
\end{equation}
where the symbol $\circ$ means outer product operation, ${\mathbf{A}_R}\in {\mathbb{C}^{{M_r} \times R}}$, ${\mathbf{A}_T}\in {\mathbb{C}^{{M_I} \times R}}$ and ${\mathbf{A}_\tau} \in {\mathbb{C}^{{N} \times R}}$ can be further expressed as
\begin{equation}
   {\mathbf{A}_R} = \left[ 
  {{{\mathbf{a}}_R}\left( {{\varphi _1}} \right)}, {{{\mathbf{a}}_R}\left( {{\varphi _2}} \right)},\cdots ,{{{\mathbf{a}}_R}\left( {{\varphi _R}} \right)} \right],
\end{equation}
\begin{equation}
    {\mathbf{A}_T} = \left[ 
  {{\mathbf{a}}_{I,T}}\left( \vartheta _{1} \right), {{\mathbf{a}}_{I,T}}\left( \vartheta _{2} \right),\cdots ,{{\mathbf{a}}_{I,T}}\left( \vartheta _{R} \right)\right],
\end{equation}
and 
\begin{equation}
    {\mathbf{A}_\tau} = \left[
    \rho_1{{{\mathbf{a}_\tau}}\left( {{\tau _1}} \right)}, \rho_2{{{\mathbf{a}_\tau}}\left( {{\tau _2}} \right)},\cdots ,\rho_R{{{\mathbf{a}}_\tau}\left( {{\tau _R}} \right)} 
 \right],
\end{equation}
where ${{\mathbf{a}}_\tau}\left( {{\tau _r}} \right)\in \mathbb{C}^{N\times 1}$ is expressed as
\begin{equation}
    {{\mathbf{a}}_{\tau}}\left( \tau_r  \right) = {\left[1,{{e^{ - j2\pi \Delta f\tau_r }}},...,{{e^{ - j2\pi \left( {N - 1} \right)\Delta f\tau_r }}}\right]^T}.
\end{equation}

Then, the CP decomposition can be achieved by solving
\begin{equation}
    \mathop {\min }\limits_{{{{\mathbf{ A}}}_R},{{{\mathbf{A}}}_T},{{{\mathbf{A}}}_\tau }} \left\| {{\boldsymbol{\mathcal{G}}_{IR}} - \sum\limits_r {{{{\mathbf{a}}}_R}\left( {{\varphi _r}} \right) \circ {{{\mathbf{a}}}_{I,T}}\left( \vartheta _{r} \right) \circ {{{\mathbf{ \tilde a}}}_\tau}\left( {{\tau _r}} \right)} } \right\|_F^2,\label{optimize}
\end{equation}
where ${{{\mathbf{ \tilde a}}}_\tau}=\rho_r{\mathbf{a}_\tau}$. To address optimization problem \eqref{optimize}, we employ the alternating least squares (ALS) algorithm, which alternately adjusts one factor matrix while maintaining the other two as constant to minimize the data fitting error until convergence \cite{tensor_Kolda,9815098}. Then the optimization problem \eqref{optimize} is further decomposed into three sub-problems, i.e., 
\begin{equation}
  {\mathbf{\hat {A}}}_{{R}}^{\left( {t } \right)} = \arg \mathop {\min }\limits_{{{{\mathbf{\hat { A}}}}_{{R}}}} \left\| {{\mathbf{{G}}}_{IR(1)}^T - \left( {{\mathbf{\hat {A}}}_{\tau}^{\left( t-1 \right)} \odot {\mathbf{\hat {A}}}_{{T}}^{\left( t-1 \right)}} \right){\mathbf{\hat {A}}}_{{R}}^T} \right\|_F^2, \label{factor_matrices_1}
\end{equation}
\begin{equation}
  {\mathbf{\hat {A}}}_{{T}}^{\left( {t } \right)} = \arg \mathop {\min }\limits_{{{{\mathbf{\hat { A}}}}_{{T}}}} \left\| {{\mathbf{{G}}}_{IR(2)}^T - \left( {{\mathbf{\hat {A}}}_{\tau}^{\left( t \right)} \odot {\mathbf{\hat {A}}}_{{R}}^{\left( t-1 \right)}} \right){\mathbf{\hat {A}}}_{{T}}^T} \right\|_F^2,\label{factor_matrices_2}
\end{equation}
\begin{equation}
  {\mathbf{\hat {A}}}_{{\tau}}^{\left( {t} \right)} = \arg \mathop {\min }\limits_{{{{\mathbf{\hat { A}}}}_{{\tau}}}} \left\| {{\mathbf{{G}}}_{IR(3)}^T - \left( {{\mathbf{\hat {A}}}_{T}^{\left( t \right)} \odot {\mathbf{\hat {A}}}_{{R}}^{\left( t \right)}} \right){\mathbf{\hat {A}}}_{{\tau}}^T} \right\|_F^2,
\label{factor_matrices_3}
\end{equation}
where $\mathbf{{G}}_{IR(1)}$, $\mathbf{{G}}_{IR(2)}$, and $\mathbf{{G}}_{IR(3)}$ are the mode-1, mode-2 and mode-3 unfolding of the tensor $\boldsymbol{\mathcal{G}}_{IR}$, respectively. $\mathbf{\hat A}^{\left(t\right)}_{R}$, $\mathbf{\hat A}^{\left(t\right)}_{T}$, and $\mathbf{\hat A}^{\left(t\right)}_{\tau}$ are the estimated matrices of factor matrices $\mathbf{A}_{R}$, $\mathbf{A}_{T}$ and $\mathbf{A}_{\tau}$ at the $t$-th iteration, accordingly. By utilizing linear minimum mean square error (LMMSE), the factors matrices containing the unknown parameters, i.e., AoA $\{\varphi_r\}_{r=1}^R$, AoD $\{\vartheta_r\}_{r=1}^R$ and delay $\{\tau_r\}_{r=1}^R$ of the targets in Eq. \eqref{factor_matrices_1} - Eq. \eqref{factor_matrices_3} can be obtained \cite{low_rank}.
{The factor matrices are initialized by the $R$ leading left singular vectors of the corresponding unfolding matrices \cite{tensor_Kolda} and the iterative process of the ALS algorithm is terminated when the relative change in the Frobenius norm of the estimated tensor between two consecutive iterations falls below a predefined threshold $\epsilon_{th}$, i.e., 
$\left\|{\boldsymbol{\mathcal{\hat G}}_{IR}^{(t)}}-{\boldsymbol{\mathcal{\hat G}}_{IR}^{(t-1)}}\right\|_F^2<\epsilon_{th}$
or the maximum number of iterations $I_{max}$ is reached. Here, ${\boldsymbol{\hat{\mathcal{G}}}}_{IR}^{(t)}$ denotes the reconstructed sensing tensor at the $t$-th ALS iteration, which is obtained from the estimated factor matrices as
\begin{equation}
\begin{split}
    {\boldsymbol{\mathcal{\hat G}}_{IR}^{(t)}} &= \sum\limits_{r = 1}^R  {{\mathbf{\hat a}}_R^{(t)}}\left( {{\varphi _r}} \right) \circ {{\mathbf{\hat a}}_{I,T}^{(t)}}\left( \vartheta _{r} \right) \circ {\mathbf{\hat {\tilde a}}^{(t)}}\left( {{\tau _r}} \right),
\end{split}\label{CPD_t}
\end{equation}
where ${{\mathbf{\hat a}}_R^{(t)}}\left( {{\varphi _r}} \right)$, ${{\mathbf{\hat a}}_{I,T}^{(t)}}\left( \vartheta _{r} \right)$, and $ {\mathbf{\hat {\tilde a}}^{(t)}}\left( {{\tau _r}} \right)$ are the $r$-th columns of the factor matrices ${\mathbf{A}_R}$, ${\mathbf{A}_T}$, and ${\mathbf{A}_\tau}$ estimated at iteration $t$, respectively.}

\subsection{Uniqueness Analysis and Parameters Estimation}
The uniqueness condition of the tensor decomposition problem is essential for estimating target parameters, as it guarantees that the resulting matrices hold accurate information about the parameters of the targets. According to Kruskal’s result \cite{Kruskal}, a condition for the uniqueness of the CP decomposition in Eq. \eqref{CPD} is as follows:
\begin{equation}
    {k_{{{{\mathbf{A}}}_R}}} + {k_{{{{\mathbf{A}}}_T}}} + {k_{{{{\mathbf{A}}}_\tau}}}\geqslant 2R+2,\label{Kruskal_uniqueness}
\end{equation}
where ${k_{\mathbf{A}_R}}$, ${k_{\mathbf{A}_T}}$ and ${k_{\mathbf{A}_{\tau}}}$ are the Kruskal rank of matrices $\mathbf{A}_R$, $\mathbf{A}_T$ and $\mathbf{A}_{\tau}$, respectively. However, the condition in \eqref{Kruskal_uniqueness} is sufficient but not necessary. Since the factor matrices have Vandermonde structure and are typically of full column rank due to the reason that different targets generally have distinct parameters, the CP decomposition is unique if 
\begin{equation}
    \min (\left({M_I}-1\right)M_r,N)  \geqslant R \label{Van_uniqueness}
\end{equation}
satisfied \cite{van_cpd}. The system usually has a limited number of targets, thus the condition in \eqref{Van_uniqueness} is generally fulfilled.

After achieving the right factor matrices, the related parameters of target $r$ can be estimated by
\begin{equation}
    {\hat \varphi }_r = \arg \mathop {\max }\limits_{\varphi}\frac{{\left| {{\bf{\hat a}}_{R,r}^H{{\bf{a}}_R}\left( \varphi  \right)} \right|}}{{{{\left\| {{{{\bf{\hat a}}}_{R,r}}} \right\|}_2}{{\left\| {{{\bf{a}}_R}\left( \varphi  \right)} \right\|}_2}}},\label{parameter_estimation_1}
\end{equation}
\begin{equation}
  {\hat \vartheta }_r = \arg \mathop {\max }\limits_{\vartheta}\frac{{\left| {{\bf{\hat a}}_{I,T,r}^H{{\bf{a}}_{I,T,r}}\left( \vartheta   \right)} \right|}}{{{{\left\| {{{{\bf{\hat a}}}_{I,T,r}}} \right\|}_2}{{\left\| {{{\bf{a}}_{I,T,r}}\left( \vartheta  \right)} \right\|}_2}}},\label{parameter_estimation_2}
\end{equation}
\begin{equation}
  {\hat \tau}_r = \arg \mathop {\max }\limits_{\tau}\frac{{\left| {{\bf{\hat  {\tilde a}}}_{\tau,r}^H{{\bf{a}}_\tau}\left( \tau  \right)} \right|}}{{{{\left\| {{{{\bf{\hat {\tilde a}}}}_{\tau,r}}} \right\|}_2}{{\left\| {{{\bf{a}}_\tau}\left( \tau  \right)} \right\|}_2}}},\label{parameter_estimation_3}
\end{equation}
where ${\bf{\hat a}}_{R,r}$, ${\bf{\hat a}}_{I,T,r}$ and ${\bf{\hat a}}_{\tau,r}$ are the $r$-th column of $\mathbf{\hat A}_R$, $\mathbf{\hat A}_{T}$ and $\mathbf{\hat A}_{\tau}$, respectively. The overall processes of the proposed CPD-based target parameter estimation scheme is summarized in \textbf{Algorithm 1}.

\begin{algorithm}[t]
    \caption{Proposed CPD-based Target Parameter Estimation Scheme}
    \justifying
    \KwIn{The received sensing signal ${\mathbf{Y}}_{\text{SEN},q,n}$ on each subcarrier $n$ and each phase configuration $q$ of RIS $I$ at sensing Rx, the channel matrix $\mathbf{G}_{TI,n}$ on each subcarrier $n$ from the Tx to the RIS $I$, the phase configuration of RIS $I$ $\mathbf{\Psi }$. }
    Obtaining the sensing channel {${{\mathbf{\hat G}}_{IR,n}}$ from RIS $I$} to the sensing Rx on each subcarrier according to Eq. \eqref{G2k} and the tensor form $\boldsymbol{\mathcal{G}}_{IR}$ according to Eq. \eqref{CPD}\;
    {Solving Eq. \eqref{optimize} by utilizing ALS for CP decomposition, which leads to Eq. \eqref{factor_matrices_1} - Eq. \eqref{factor_matrices_3} to iteratively estimate the factor matrices $\mathbf{\hat A}_R$, $\mathbf{\hat A}_T$ and $\mathbf{\hat A}_\tau$ until convergence}\;
    \textbf{For} {$r=1:R$} \textbf{do}\\
    Estimate the target parameters AoA ${{\hat \varphi }_r}$, AoD ${\hat \vartheta}_{r}$ and delay $\hat{\tau}_r$ according to Eq. \eqref{parameter_estimation_1} - Eq. \eqref{parameter_estimation_3}\;
    \textbf{end}\\
    
    \vspace{-1em}
    \KwOut{Estimated AoA ${{\hat \varphi }_r}$, AoD ${\hat \vartheta}_{r}$, and delay $\hat{\tau}_r$ of the targets.}
\end{algorithm}

Then the position ${\mathbf{u}_r}\in \mathbb{R}^{1\times 2}$ of the target $r$ can be estimated directly by using the estimated AoA ${{\hat \varphi }_r}$, AoD ${\hat \vartheta }_{r}$ and delay $\hat{\tau}_r$. Specifically, assume the position of the RIS $I$ is denoted as $\mathbf{p}_I \in \mathbb{R}^{1\times 2}$ and the position of the sensing Rx is denoted as $\mathbf{p}_R \in \mathbb{R}^{1\times 2}$, then the estimated position $\hat {\mathbf{u}}_r$ can be calculated by
\begin{equation}
    \hat {\mathbf{u}}_r= {\mathbf{p}}_I+\left(c\hat{\tau}_r-\hat{d}_r\right)\mathbf{f}_r,\label{ur}
\end{equation}
where $\hat{d}_r$ is the estimated distance from target $r$ to the sensing receiver and $\mathbf{f}_r=\left[\sin{{\hat\vartheta }_{r}},\cos{{{\hat \vartheta}_{r}}}\right]$. And $\hat{d}_r$ can be calculated by
\begin{equation}
\left[ {\begin{array}{*{20}{c}}
  {{{\mathbf{g}}_1^T}}& \cdots &{{{\mathbf{0}}_{2 \times 1}}} \\  
   \vdots &  \ddots & \vdots  \\ 
  {{{\mathbf{0}}_{2 \times 1}}}& \cdots &{{{\mathbf{g}}_R^T}} 
\end{array}} \right]\left[ {\begin{array}{*{20}{c}}
  {{d_1}} \\ 
   \vdots  \\ 
  {{d_R}} 
\end{array}} \right]\!\!\! = \!\!\! \left[ {\begin{array}{*{20}{c}}
  { c{{\hat \tau }_1}{\mathbf{f}}_1^T+{\mathbf{p}}_I^T -{\mathbf{p}}_R^T} \\ 
   \vdots  \\ 
  { c{{\hat \tau }_R}{\mathbf{f}}_R^T+{\mathbf{p}}_I^T -{\mathbf{p}}_R^T} 
\end{array}} \right],
\end{equation}
where $\mathbf{g}_r=\left[\sin{\hat{\varphi} _r}+\sin{\hat{\vartheta} _r},-\cos{\hat{\varphi} _r}+\cos{\hat{\vartheta} _r}\right]$.

\textcolor{black}{To directly characterize the localization accuracy of the proposed sensing target parameter estimation scheme, the RMSE is employed and is expressed as follows \cite{9815098}}
\begin{equation} \label{RMSE}
    {\text{RMSE}} = \sqrt { {\frac{1}{R}{\sum\limits_{r = 1}^R{\left\| {\hat {\mathbf{u}}_r - {\mathbf{u}_r}} \right\|}^2}} },
\end{equation}
where $\hat {\mathbf{u}}_r$ is the estimated position of the $r$-th target and $\mathbf{u}_r$ is the corresponding true location. Until now, we have introduced detailed processes about how to extract the location parameters from the received signals at the sensing Rx. Since our goal is to minimize the sensing error while satisfying the communication requirements, in the next section, we will discuss how to formulate and solve the optimization problem based on the system requirements.

\section{Joint Beamforming and Phase Shift Design for Considered ISAC System }
Up to this point, we have introduced the communication objectives and the corresponding sensing procedures. In the considered ISAC system, a fundamental trade-off exists between communication performance and sensing accuracy. Therefore, the beamforming design at the transmitter and the phase shift optimization at the RISs must be carefully orchestrated to meet the demands of both functionalities. In this section, we formulate an optimization problem to maximize sensing accuracy while satisfying the communication requirements. The detailed formulation is presented below.

\subsection{Problem Formulation for Considered ISAC Systems}
In this paper, our goal of the ISAC design is that we want to minimize the RMSE of the sensing results while guaranteeing the SNR requirements for the communication users. Based on that, the overall optimization problem can be formulated as
	\begin{align}
			\textbf{(P0)}  &\min_{\boldsymbol{w}_{k,j},\mathbf{\Phi}_{i}} \,\,\, \text{RMSE}, \label{obj11}	\\ 
			&  \,\,\,\,\,\,\,\, \,\, s.t. \,\,\,\,\,\,\,\,\,\,\, \gamma_{k,j} \geq \gamma_{thr}, \,\,k=1,2,...,K,\,\,j=1,2,...,J, \label{C11} \\ 
			& \,\,\,\,\,\,\,\, \,\, \,\,\,\,\,\,\,\,\,\,\,\, \,\,\,\,\,\,\,\sum_{k=1}^{K}\sum_{j=1}^{J} ||\boldsymbol{w}_{k,j}||^2 \leq P_{max},  \label{C21}\\
            & \,\,\,\,\,\,\,\, \,\, \,\,\,\,\,\,\,\,\,\,\,\, \,\,\,\,\,\,\, \textcolor{black}{|\phi_{i}(m)|=1, \,\,\, i=1,2,...,I, \,\, m=1,2,...,M_i.} \label{C31}
	\end{align}

The optimization variables include the beamforming vector design $\boldsymbol{w}_{k,j}$ at the Tx side, \textcolor{black}{the phase shift matrices $\mathbf{\Phi}_i$ of each RIS $i$.} \textcolor{black}{ The constraint \eqref{C11} represents the SNR requirements for each communication user, the value of this threshold $\gamma_{thr}$ reflects the importance we place on communication, a higher threshold indicates greater emphasis on communication, and we optimize sensing while ensuring communication requirements are met, which guarantees a balanced design for both communication and sensing.} The constraint \eqref{C21} is the maximum transmit power constraint at the the Tx side and constraint \eqref{C31} is the unit power constraint. In the following section, we will give detailed instructions about how to solve this optimization problem. 

First, for the objective function expressed in the optimization problem \eqref{obj11}, the location estimation results strongly depend on the received signal strength at the sensing Rx side. Generally, higher received signal strength leads to improved estimation accuracy. This relationship has been widely established in wireless localization literature \cite{Power1,Power2}.
 Based on this fact, we change the original objective function, i.e., minimize RMSE of the estimation results, into maximizing the received signal strength that reflected by the $R$ targets at the sensing Rx side. As we defined before, the whole sensing channel from Tx to sensing Rx can be expressed as shown in Eq. \eqref{sensing_channel}. The sensing Rx will reuse all the communication signals for all $K$ users that are reflected by $R$ sensing targets to extract the location information of them. As discussed in Section \ref{Sen_Model}, for sensing purposes, an initial scanning phase is required, during which $Q$ beam patterns are generated to perform wide-beam scanning in order to identify the approximate directions of the targets. Once these coarse directions are determined, narrow and high-gain beams are subsequently designed and steered toward these directions to illuminate the targets, thereby enabling precise location estimation. In the considered system, our objective is to enhance sensing accuracy to the highest possible level in order to ensure the safety of home monitoring. To achieve this, we focus on maximizing the received sensing signal strength at the sensing Rx. Thereby in the following optimization process, we ignore the index $q$, $q=1,2,...,Q$, which means that we always want to maximize the received signal strength at the sensing Rx. By neglecting the index $q$, the received signal on subcarrier $j$ of user $k$ at the sensing Rx can be expressed as
\begin{equation}
\boldsymbol{y}_{\text{SEN},k,j}=\mathbf{H}_{\text{SEN},k,j}\boldsymbol{w}_{k,j}x_{k,j} + \boldsymbol{n}_{k,j},
\end{equation}
the channel $\mathbf{H}_{\text{SEN},k,j}$ from Tx to sensing Rx on the $j$-th subcarrier of user $k$ is generated in the same way as we describe in Eq. \eqref{sensing_channel} by ignoring the index $q$. And $\boldsymbol{n}_{k,j}\in \mathbb{C}^{M_r \times 1}$ represents the noise at the sensing Rx side, and each element of it follows the complex Gaussian distribution with zero mean and variance $\sigma_s^2$. Thereby the overall received signal strength at the sensing Rx can be expressed as

\begin{equation}
P_{\text{SEN}}=\sum_{k=1}^{K}\sum_{j=1}^{J}||\boldsymbol{y}_{\text{SEN},k,j}||^2.
\end{equation}
With the above definition, we can transform the original objective problem into the following
	\begin{align}
			\textbf{(P1)}  &\max_{\boldsymbol{w}_{k,j},\mathbf{\Phi}_{i}} \,\,\,\, P_{\text{SEN}},	\\ 
			&  \,\,\,\,\,\,\,\, \,\, s.t. \,\,\,\,\,\,\,\,\,\,\,\, \eqref{C11},\eqref{C21},\eqref{C31}. \nonumber 
	\end{align}

The above problem is still hard to solve since the optimization variables $\boldsymbol{w}_{k,j}$ and \textcolor{black}{$\mathbf{\Phi}_i$} are coupled together. To solve this problem, we decouple $\textbf{(P1)} $ into three sub-problems, the first sub-problem is used to optimize the beamforming design at the Tx, \textcolor{black}{the second sub-problem is used to update the phase shift designs at RIS $i$, $i=1,2,..., I$, respectively.} In the following, we will describe each sub-problem in detail.

\subsection{Beamforming Design at the Tx}
First, we want to optimize the beamforming vectors $\boldsymbol{w}_{k,j}$ for each user $k$ on subcarrier $j$ at the Tx side, while the phase shift matrices \textcolor{black}{ $\mathbf{\Phi}_i$, $i=1,2,...,I$} are fixed. This sub-problem can be expressed as 
	\begin{align}
			\textbf{(SP-1.0)} \,\, &\max_{\boldsymbol{w}_{k,j}} \,\,\,\, \,\,\,\,\,\,\,\,\,\,\,\,\, P_{\text{SEN}},	\\ 
			&   \,\,\, s.t. \,\,\,\,\,\,\,\,\,\,\,\,\,\,\,\, \gamma_{k,j} \geq \gamma_{\text{thr}}, \label{SP111} \\ 
			& \,\,\,\,\,\,\,\, \,\, \,\,\,\,\,\,\,\,\,\,\,\, \,\,\,\,\,\,\,\, \sum_{k=1}^{K} \sum_{j=1}^{J}||\boldsymbol{w}_{k,j}||^2 \leq P_{\text{max}}. \label{SP121}
	\end{align}

To solve this problem, we first introduce an auxiliary matrices $\mathbf{W}_{k,j}=\boldsymbol{w}_{k,j}\boldsymbol{w}_{k,j}^H \in \mathbb{C}^{M_t \times M_t}$ for each user $k$ on subcarrier $j$, which is a hermitian semi-definite matrix. Then $\textbf{(SP-1.0)}$ can be shown as 
\begin{align}
			\textbf{(SP-1.1)}\,\,  &\max_{\boldsymbol{w}_{k,j}} \,\,\,\, \,\,\,\,\,\,\,\,\,\,\,\,\, P_{\text{SEN}}, \label{111}	\\ 
			&   \,\,\,\,\,\,\,\, \,\, s.t. \,\,\,\,\,\,\,\,\,\,\,\,  \boldsymbol{h}_{k,j}\mathbf{W}_{k,j} \boldsymbol{h}^H_{k,j}\geq \gamma_{\text{thr}}\sigma_c^2, \label{SP211_1} \\ 
			& \,\,\,\,\,\,\,\, \,\, \,\,\,\,\,\,\,\,\,\,\,\, \,\,\,\,\,\,\,\,\sum_{k=1}^K \sum_{j=1}^J\mathrm{Tr}\left( \mathbf{W}_{k,j} \right) \leq P_{\text{max}}, \label{SP221_1}
            \\& \,\,\,\,\,\,\,\, \,\, \,\,\,\,\,\,\,\,\,\,\,\, \,\,\,\,\,\,\,\, \mathbf{W}_{k,j} \geq \mathbf{0}.\label{SP231}
	\end{align}
By introducing the auxiliary matrices $\mathbf{W}_{k,j}$, we successfully transform \textbf{(SP-1.1)} as a Quadratically Constrained Quadratic Programming (QCQP) problem, which is 
convex and ready to be solved by convex optimization. We assume that the obtained optimal solution for $\mathbf{W}_{k,j}$ is $\mathbf{W}_{k,j}^{*}$, then we need to recover the original beamforming vector $\boldsymbol{w}_{k,j}^{*}$ from $\mathbf{W}_{k,j}^{*}$. First, we apply the singular value decomposition (SVD) of $\mathbf{W}_{k,j}^{*}$ as 
\begin{equation} \label{SVD1}
    \left[\mathbf{U}_0, \mathbf{\Sigma}_0, \mathbf{V}_0   \right] =\text{SVD}\left(\mathbf{W}_{k,j}^{*}\right),
\end{equation}
then we use the eigenvector with the largest singular value as the  final result for $\boldsymbol{w}_{k,j}^{*}$ \cite{SDR}, i.e.,
\begin{equation}\label{SVD2}
    \boldsymbol{w}_{k,j}^{*}=\mathbf{U}_0(:,1)\sqrt{\mathbf{\Sigma}_0(1,1)}
\end{equation}
Based on the above process, we can optimize the beamforming design at the Tx side. 

\subsection{\textcolor{black}{Sequential Phase Shift Optimization at RISs}}
\textcolor{black}{
In this subsection, we optimize the phase shifts of multiple RISs in a sequential manner.
Specifically, when optimizing the phase shift vector of RIS $i$, denoted by $\boldsymbol{\phi}_i$, the Tx beamforming vectors and the phase shift vectors of all
other RISs, i.e., $\{\boldsymbol{\phi}_{\ell}\}_{\ell\in\mathcal{I},\ell\neq i}$, are fixed. Then, the phase shift vectors of different RISs are updated one by one. For each RIS $i\in\mathcal{I}$, we define the set of communication users whose serving
paths include RIS $i$ as
\begin{equation}
    \mathbb{K}_i
    \triangleq
    \left\{
    k\in\mathbb{K}_{\rm RIS}: i\in\mathcal{P}_k
    \right\}.
\end{equation}
For user $k\in\mathbb{K}_i$, we denote by $\ell_{k,i}$ the position of RIS $i$ in the
ordered serving path $\mathcal{P}_k$, i.e.,
\begin{equation}
    i_{k,\ell_{k,i}}=i.
\end{equation}
When optimizing RIS $i$, only the users in $\mathbb{K}_i$ are affected by
$\boldsymbol{\phi}_i$. Therefore, the communication constraints of users not passing
through RIS $i$ can be omitted in this subproblem since they remain unchanged. With fixed Tx beamforming vectors and fixed phase shifts of all other RISs, the phase
shift optimization subproblem for RIS $i$ can be formulated as
\begin{align}
\textbf{(SP-2.0)} \quad
& \max_{\boldsymbol{\phi}_i}
\quad
P_{\rm SEN} \\
& \mathrm{s.t.}
\quad
\gamma_{k,j}\geq \gamma_{\rm thr},
\quad
k\in\mathbb{K}_i,\ j=1,2,\ldots,J_k, \label{SP_RIS_i_SNR}\\
& \phantom{\mathrm{s.t.}}
\quad
|\phi_i(m)|=1,
\quad
m=1,2,\ldots,M_i. \label{SP_RIS_i_unit}
\end{align}}

\textcolor{black}{
To solve \textbf{(SP-2.0)}, we first reformulate the sensing and communication
channels by extracting the phase shift vector $\boldsymbol{\phi}_i$ from the cascaded
channel. For the sensing link on subcarrier $n$, the equivalent channel from the Tx to
the sensing Rx can be rewritten as
\begin{equation}
    \boldsymbol{g}_{TR,n}
    =
    \mathbf{A}^{\rm sen}_{i,n}
    \mathbf{\Phi}_i
    \boldsymbol{b}^{\rm sen}_{i,n},
\end{equation}
where $\boldsymbol{b}^{\rm sen}_{i,n}\in\mathbb{C}^{M_i\times 1}$ denotes the incident
signal arriving at RIS $i$ from the Tx side, and
$\mathbf{A}^{\rm sen}_{i,n}$ denotes the equivalent sensing channel from RIS $i$ to the
sensing Rx, with all other RIS phase shift matrices fixed. Since
$\mathbf{\Phi}_i=\mathrm{diag}(\boldsymbol{\phi}_i)$, we have
\begin{align}
    \boldsymbol{g}_{TR,n}
    &=
    \mathbf{A}^{\rm sen}_{i,n}
    \mathrm{diag}\left(\boldsymbol{b}^{\rm sen}_{i,n}\right)
    \boldsymbol{\phi}_i \notag\\
    &=
    \mathbf{G}^{\rm sen}_{i,n}\boldsymbol{\phi}_i,
    \label{sensing_reform_general}
\end{align}
where
\begin{equation}
    \mathbf{G}^{\rm sen}_{i,n}
   =
    \mathbf{A}^{\rm sen}_{i,n}
    \mathrm{diag}\left(\boldsymbol{b}^{\rm sen}_{i,n}\right).
\end{equation}}

\textcolor{black}{
Similarly, for each communication user $k\in\mathbb{K}_i$, the equivalent communication
channel after Tx beamforming can be reformulated as
\begin{equation}
    \boldsymbol{h}_{k,j}\boldsymbol{w}_{k,j}
    =
    \boldsymbol{a}_{i,k,j}
    \mathbf{\Phi}_i
    \boldsymbol{b}_{i,k,j},
\end{equation}
where $\boldsymbol{b}_{i,k,j}\in\mathbb{C}^{M_i\times 1}$ denotes the equivalent incident
signal arriving at RIS $i$ from the Tx side, and
$\boldsymbol{a}_{i,k,j}\in\mathbb{C}^{1\times M_i}$ denotes the equivalent channel from
RIS $i$ to user $k$, including all subsequent RISs on the serving path of user $k$.
Therefore,
\begin{align}
    \boldsymbol{h}_{k,j}\boldsymbol{w}_{k,j}
    &=
    \boldsymbol{a}_{i,k,j}
    \mathrm{diag}\left(\boldsymbol{b}_{i,k,j}\right)
    \boldsymbol{\phi}_i \notag\\
    &=
    \boldsymbol{h}_{i,k,j}\boldsymbol{\phi}_i,
    \label{comm_reform_general}
\end{align}
where
\begin{equation}
    \boldsymbol{h}_{i,k,j}
   =
    \boldsymbol{a}_{i,k,j}
    \mathrm{diag}\left(\boldsymbol{b}_{i,k,j}\right)
    \in\mathbb{C}^{1\times M_i}.
\end{equation}}

\textcolor{black}{
With the above reformulated channels, the sensing signal power can be expressed as
\begin{equation}
    P_{\rm SEN}
    =
    \sum_{n=1}^{N}
    \left\|
    \mathbf{G}^{\rm sen}_{i,n}\boldsymbol{\phi}_i
    \right\|_2^2.
\end{equation}
Then, by defining the auxiliary variable
\begin{equation}
    \mathbf{\Xi}_i
    =
    \boldsymbol{\phi}_i\boldsymbol{\phi}_i^H
    \in\mathbb{C}^{M_i\times M_i},
\end{equation}
the phase shift optimization problem for RIS $i$ can be reformulated as
\begin{align}
\textbf{(SP-2.1)} \quad
& \max_{\mathbf{\Xi}_i}
\quad
\sum_{n=1}^{N}
\mathrm{Tr}
\left(
\mathbf{G}^{\rm sen}_{i,n}
\mathbf{\Xi}_i
\left(\mathbf{G}^{\rm sen}_{i,n}\right)^H
\right)
\label{SP_RIS_i_obj}\\
& \mathrm{s.t.}
\quad
\boldsymbol{h}_{i,k,j}
\mathbf{\Xi}_i
\boldsymbol{h}_{i,k,j}^{H}
\!\geq\!
\gamma_{\rm thr}\sigma_c^2,
\,
k\!\in\!\mathbb{K}_i,\,j\!\!=\!\!1,2,\ldots,J,
\label{SP_RIS_i_SNR_SDR}\\
& \phantom{\mathrm{s.t.}}
\quad
\mathrm{diag}(\mathbf{\Xi}_i)=\mathbf{1}_{M_i},
\label{SP_RIS_i_diag}\\
& \phantom{\mathrm{s.t.}}
\quad
\mathbf{\Xi}_i\succeq \mathbf{0}.
\label{SP_RIS_i_PSD}
\end{align}
The original rank-one constraint $\mathrm{rank}(\mathbf{\Xi}_i)=1$ is omitted to obtain
a convex semidefinite relaxation (SDR) problem, which can be efficiently solved by CVX. And this rank-one constraint will be recovered by the following Gaussian randomization process.}

\textcolor{black}{
After obtaining the optimal solution $\mathbf{\Xi}_i^*$, we recover the phase shift
vector $\boldsymbol{\phi}_i$ using Gaussian randomization \cite{Guassian_Random}. Specifically, we first perform
the eigenvalue decomposition
\begin{equation} \label{rec_RIS0}
    \mathbf{\Xi}_i^*
    =   \mathbf{U}_i\mathbf{S}_i\mathbf{U}_i^H.
\end{equation}
Then, we generate $N_{\rm rand}$ random vectors
$\boldsymbol{r}_{q}\sim\mathcal{CN}(\mathbf{0},\mathbf{I})$,
$q=1,2,\ldots,N_{\rm rand}$, and construct
\begin{equation} \label{rec_RIS1}
\tilde{\boldsymbol{\phi}}_{i,q}
    =    \mathbf{U}_i\mathbf{S}_i^{1/2}\boldsymbol{r}_{q}.
\end{equation}
Finally, we select the phase shift vector which has the highest sensing power $\sum_{n=1}^{N_{\rm rand}}\mathrm{Tr}\left(\mathbf{G}^{\rm sen}_{i,n}\tilde{\boldsymbol{\phi}}_{i,q}\tilde{\boldsymbol{\phi}}_{i,q}^H (\mathbf{G}^{\rm sen}_{i,n})^H\right)$ as the final update result, i.e.,
\begin{equation} \label{45}
\tilde{\boldsymbol{\phi}}_{i}^*=\text{argmax}_{\tilde{\boldsymbol{\phi}}_{i,q}}\sum_{n=1}^{N_{\rm rand}} \mathrm{Tr}\left(\mathbf{G}^{\rm sen}_{i,n}\tilde{\boldsymbol{\phi}}_{i,q}\tilde{\boldsymbol{\phi}}_{i,q}^H (\mathbf{G}^{\rm sen}_{i,n})^H\right).
\end{equation}
 It has been shown that such an semidefinite relaxation (SDR) approach followed by sufficiently large
number of randomizations of $\boldsymbol{\phi}_{i,q}$ guarantees an $\frac{\pi}{4}$-approximation
of the optimal objective value of \textbf{(SP-2.1)}  \cite{gap}.}

\begin{algorithm}[t!] 
    \caption{Interactive Optimization to Maximize the Received Sensing Signal Strength at Sensing Rx} 
    \justifying
    \KwIn{The channel $\boldsymbol{h}_{k,j}$ for communication user $k$ on subcarrier $j$; 
    the sensing channel $\mathbf{H}_{\mathrm{SEN},n}$ on each subcarrier $n$ from the Tx to 
    sensing Rx; the maximum available transmit power $P_{\max}$ at the Tx; the SNR threshold 
    $\gamma_{\mathrm{thr}}$ for communication users; the variances $\sigma_c^2$ and 
    $\sigma_s^2$ of the noise; the maximum iteration time $T$.}

    \textbf{For} {$t=1:T$} \textbf{do}\\
    Formulate subproblem \textbf{(SP-1.1)} for Tx beamforming optimization with fixed RIS 
    phase shift vectors $\{\boldsymbol{\phi}_i\}_{i\in\mathcal{I}}$\;
    
    Solve subproblem \textbf{(SP-1.1)} using CVX optimization toolbox, then recover the 
    optimal beamforming vectors $\{\boldsymbol{w}_{k,j}^{*}\}$ based on Eq. \eqref{SVD1} and Eq. \eqref{SVD2} \;
    
    \textcolor{black}{\textbf{For each} RIS $i\in\mathcal{I}$ \textbf{do}}\\
     \textcolor{black}{Fix the Tx beamforming vectors $\{\boldsymbol{w}_{k,j}^{*}\}$ and the phase shift 
    vectors of all other RISs, i.e., 
    $\{\boldsymbol{\phi}_{\ell}\}_{\ell\in\mathcal{I},\ell\neq i}$\;}
    
    \textcolor{black}{Reformulate the sensing channel $\boldsymbol{g}_{TR,n}$ by extracting 
    $\boldsymbol{\phi}_i$ according to Eq.~\eqref{sensing_reform_general}\;}
    
    \textcolor{black}{Reformulate the communication channels for users 
    $k\in\mathbb{K}_i$ by extracting $\boldsymbol{\phi}_i$ according to Eq.~\eqref{comm_reform_general}\;}
    
    \textcolor{black}{Formulate subproblem \textbf{(SP-2.1)} according to 
    Eq.~\eqref{SP_RIS_i_obj} - Eq.~\eqref{SP_RIS_i_PSD}\;}
    
    \textcolor{black}{Solve subproblem \textbf{(SP-2.1)} using CVX optimization toolbox, then recover 
    the optimal phase shift vector $\boldsymbol{\phi}_i^{*}$ through Gaussian randomization based on Eq. \eqref{rec_RIS0} - Eq. \eqref{45}\;}
    
    Update $\boldsymbol{\phi}_i\leftarrow\boldsymbol{\phi}_i^{*}$\;
    \textbf{end}\\

    \vspace{-1em}
    \KwOut{The optimal beamforming vectors $\{\boldsymbol{w}_{k,j}^{*}\}$, the optimal 
    phase shift vectors $\{\boldsymbol{\phi}_i^{*}\}_{i\in\mathcal{I}}$, and the optimal 
    received sensing signal strength $P_{\mathrm{SEN}}^{*}$ at the sensing Rx.}
\end{algorithm}

\begin{algorithm}[t!] 
    \caption{Overall Processes of the Considered ISAC System for Home Security Applications  } 
    \justifying

\KwIn{The channel $\boldsymbol{h}_{k,j}$ for communication user $k$ on subcarrier $j$; the channel $\mathbf{H}_{\text{SEN},n}$ on each subcarrier $n$ from the Tx to sensing Rx; the maximum available transmit power $P_{\text{max}}$ at the Tx, the SNR threshold $\gamma_{\text{thr}}$ for communication user, the variances $\sigma_c^2$ and $\sigma_s^2$ of the noise. }

\textbf{Beam Scanning:} RIS $I$ generate $Q$ wide beam to sweep the target area, sensing Rx receives weak sensing signals and leverages Algorithm 1 to get the rough number and directions of potential targets\;

\textbf{Sensing Signal Optimization:} Optimize the system design according to Algorithm 2 and obtain the highest sensing signal strength $P_{\text{sen}}^*$  \;

\textbf{Precise Location Extraction:} Estimate the precise location parameters. e.g., AoAs, AoDs, and delays of $R$ sensing targets according to Algorithm 1\;

\KwOut{Estimated accurate AoA ${{\hat \varphi }_r}$, AoD ${{\hat \vartheta }_{r}}$, and delay $\hat{\tau}_r$ of each target. Calculate locations of targets based on {Eq. \eqref{ur}} and the final RMSE according to Eq. \eqref{RMSE}.}
\end{algorithm}


Up to now, we have completed all the optimizations for Tx and RISs. The overall optimization process is summarized in \textbf{Algorithm 2}. With the optimal $\boldsymbol{w}_{k,j}^*$ and $\mathbf{\Phi}_i^*$, the sensing Rx can obtain the highest received signal power from all the subcarriers to performance localization tasks. The overall design for the considered ISAC system is summarized in  \textbf{Algorithm 3}.
In Algorithm 3, the first step is beam scanning phase. As we introduce before, in this period, wide beams are generated to sweep the whole target area for obtaining preliminary information about sensing targets. With rough number and directions of these sensing targets, we then generate narrow and strong beams to illuminate these target for extracting precise locations information, e.g., AoA ${{\hat \varphi }_r}$, AoD ${{\hat \nu }_{2,T,r}}$, and delay $\hat{\tau}_r$ of the target $r$, $r=1,2,...,R$. Then we calculate the 
2D location of these targets based on {Eq. \eqref{ur}} and the final RMSE according to Eq. \eqref{RMSE}. 

\section{Performance Analysis}
In this section, we provide detailed complexity and CRLB analyses about the proposed scheme for the home security application in this work. We first start with the complexity analysis, then we provide the CRLB analyses to demonstrate that the performance of our proposed scheme can approach the theoretical optimal performance.

\subsection{Complexity Analysis}
The computational complexity of our proposed \textcolor{black}{multi-RIS assisted} target parameter estimation scheme contains both the complexity of parameter estimation in Algorithm 1 and beamforming optimization in Algorithm 2. The complexity of Algorithm 1 is presented as follows. Specifically, the complexity of obtaining the sensing channel tensor $\boldsymbol{\mathcal{G}}_{IR}$ in Algorithm 1 is $\mathcal{O}\left(M_IM_rM_tQN\right)$. 
The complexity of the tensor decomposition-based parameter estimation method is dominated by the ALS for factor matrices estimation and is of order $\mathcal{O}\left(I_{ite}M_IM_rNR\right)$ \cite{low_rank}, where $I_{ite}$ is the number of iterations for ALS. 
For the optimization problem proposed in Algorithm 2, the complexity for obtaining the optimal beamforming vector $\boldsymbol{w}_k^*$ is $\mathcal{O}\left( M_t^{3.5}\right)$ \cite{SDR}; \textcolor{black}{For the phase shifts update at RISs, the complexity of getting the optimal auxiliary variable $\mathbf{\Xi}_i^*$ in Eq. \eqref{rec_RIS0} is $\mathcal{O}\left( M_i^{3.5}\right)$ \cite{SDR}. Combining all the above, the overall complexity for Algorithm 2 is $\mathcal{O}\left(K M_t^{3.5}+ \sum_{i=1}^{I}M_i^{3.5} \right)$.
We summarized the whole ISAC processes in Algorithm 3, which is a combination of Algorithm 1 and Algorithm 2, thereby the final complexity for the complete system design is $\mathcal{O}\big(M_IM_rN(M_tQ+RI_{ite})+ KM_t^{3.5}+\sum_{i=1}^{I} M_i^{3.5}\big)$.}

\subsection{Cramér-Rao Lower Bound Analysis} 
The CRLB is a lower bound on the variance of the unbiased estimator and also serves as a benchmark for evaluating the performance of the proposed scheme \cite{942635}. It can be expressed as the inverse of  the Fisher information matrix (FIM) about the unkown parameters, i.e.,
\begin{equation}    {\text{CRLB}}\left(\boldsymbol{\Theta}\right) = {\boldsymbol{\Omega} ^{ - 1}}\left( \boldsymbol{\Theta} \right),
\end{equation}
where $\boldsymbol{\Theta} = \left[{\boldsymbol{\varphi }}, {\boldsymbol{\vartheta}}, \boldsymbol{\tau} \right]$, with ${\boldsymbol{\varphi }}=\left[\varphi_1,\varphi_2,...,\varphi_R\right]$, ${\boldsymbol{\vartheta}}=\left[{\vartheta}_{1},{\vartheta}_{2},...,{\vartheta}_{R}\right]$ and $\boldsymbol{\tau}=\left[\tau_1,\tau_2,...,\tau_R\right]$. Accordingly, the log-likelihood function of $\boldsymbol{\Theta}$ has the expression of 
\begin{small}
\begin{equation}
\begin{split}
  L\left( \boldsymbol{\Theta} \right) &= f\left( {\boldsymbol{\mathcal{G}_{IR}};\mathbf{A}_R,\mathbf{A}_T,\mathbf{A}_{\tau}} \right)\\
   &=  - M_rM_IN\ln \left( {\pi {\sigma_s ^2}} \right) - \frac{1}{{{\sigma_s ^2}}}\left\| {{\mathbf{G}}_{IR(1)}^T - \left( {\mathbf{A}_{\tau} \odot \mathbf{A}_T} \right){\mathbf{A}_R^T}} \right\|_F^2 \\
   &=  - M_rM_IN\ln \left( {\pi {\sigma_s ^2}} \right) - \frac{1}{{{\sigma_s ^2}}}\left\| {{\mathbf{G}}_{IR(2)}^T - \left( {\mathbf{A}_{\tau} \odot \mathbf{A}_R} \right){\mathbf{A}_T^T}} \right\|_F^2 \\
   &=  - M_rM_IN\ln \left( {\pi {\sigma_s ^2}} \right) - \frac{1}{{{\sigma_s ^2}}}\left\| {{\mathbf{G}}_{IR(3)}^T - \left( {\mathbf{A}_T \odot \mathbf{A}_R} \right){\mathbf{A}_{\tau}^T}} \right\|_F^2.
\end{split}  
\end{equation}
\end{small}

Then, the FIM for $\boldsymbol{\Theta}$ can be written as 
\begin{equation}
    \boldsymbol{\Omega} \left( \boldsymbol{\Theta} \right) = \mathbb{E}\left\{ {{{\left( {\frac{{\partial L\left( \boldsymbol{\Theta} \right)}}{{\partial \boldsymbol{\Theta}}}} \right)}^H}\left( {\frac{{\partial L\left( \boldsymbol{\Theta}\right)}}{{\partial \boldsymbol{\Theta}}}} \right)} \right\}.\label{Omega}
\end{equation}

More details about how to derive Eq. \eqref{Omega} can be found in reference \cite{low_rank}. 

In the next section, we will provide simulation results of our proposed parameter estimation scheme compared with the corresponding CRLB to demonstrate that the performance of our proposed scheme can approach the theoretical optimal performance under certain transmit power conditions, and then provide comparisons of the localization RMSE performance of our proposed scheme with that of state-of-the-art techniques to validate the superiority of our proposed method.
\begin{figure}
\centering
\includegraphics[width=\linewidth]{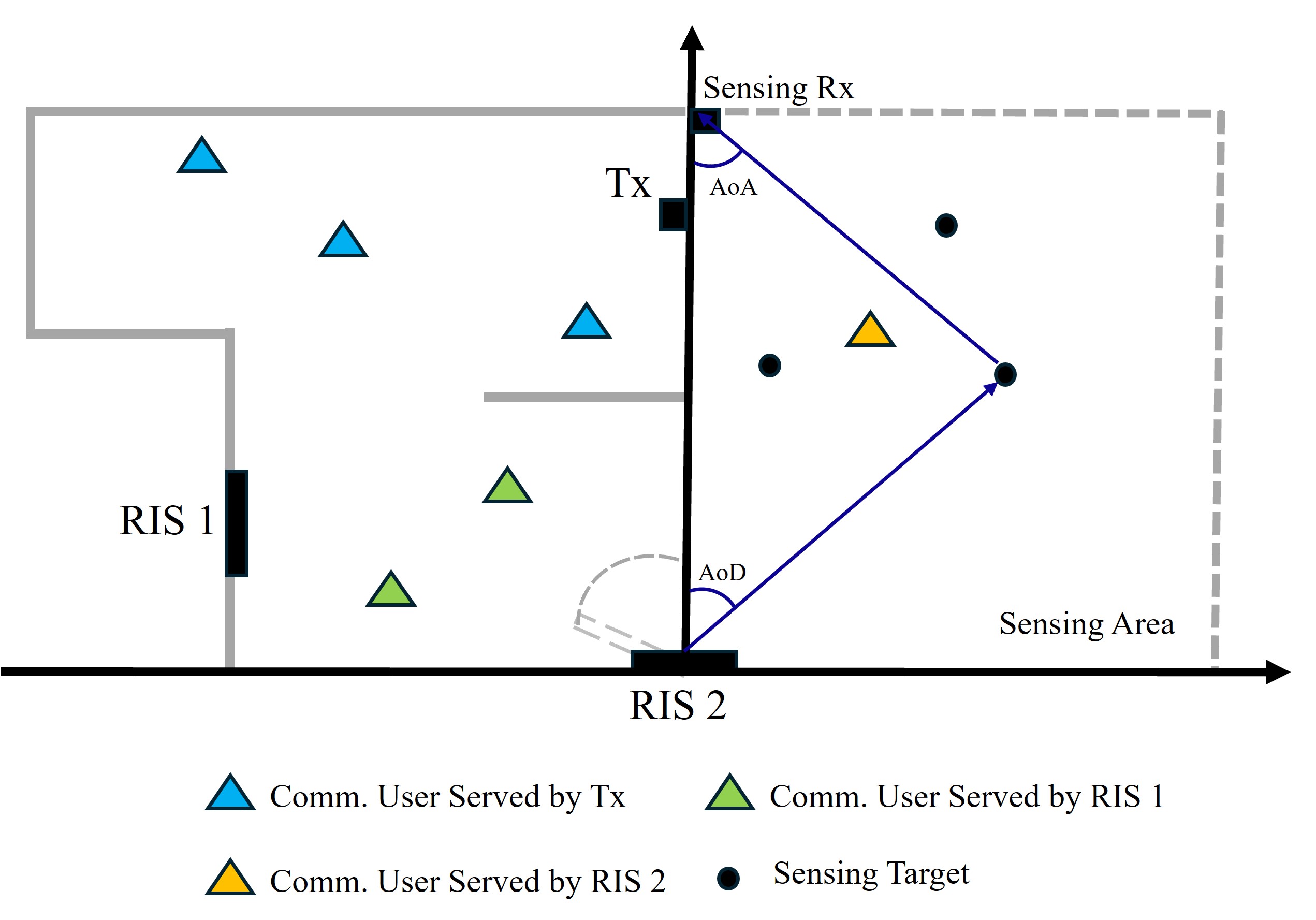}
\caption{Geometric illustration of the system. The total number of communication users is $K=K_0+K_1+K_2=6$. The number of sensing targets is $R=3$. The sensing area for the sensing Rx is set to be $20\, \mathrm{m} \times 20\,\mathrm{m}$. The AoAs and AoDs are the corresponding values from $0$ to $\pi/2$ since the target area is only in the right half of RIS $2$'s service area.}
\label{simulation_setting}
\end{figure}

\begin{figure}[t]
\centering
\includegraphics[width=\linewidth]{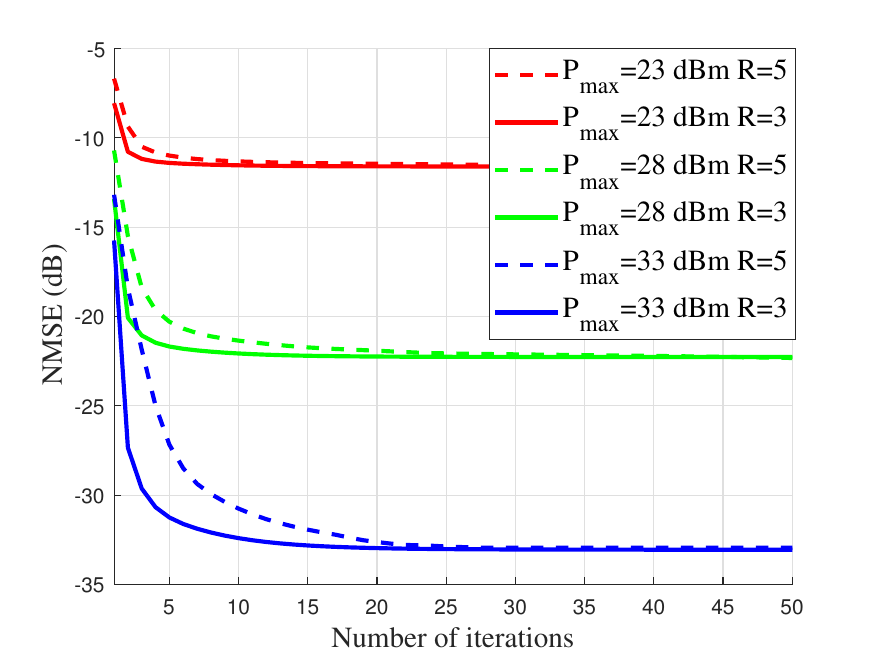}
\caption{\textcolor{black}{NMSE of $ {\boldsymbol{\mathcal{G}}_{IR}}$ versus the number of iterations under different transmit powers and target numbers.}}
\label{iteration}
\end{figure}

\begin{figure*}
	\setlength{\abovecaptionskip}{-5pt}
	\setlength{\belowcaptionskip}{-10pt}
	\centering
	\begin{minipage}[b]{0.33\linewidth}
		\centering
		\includegraphics[width=2.3in]{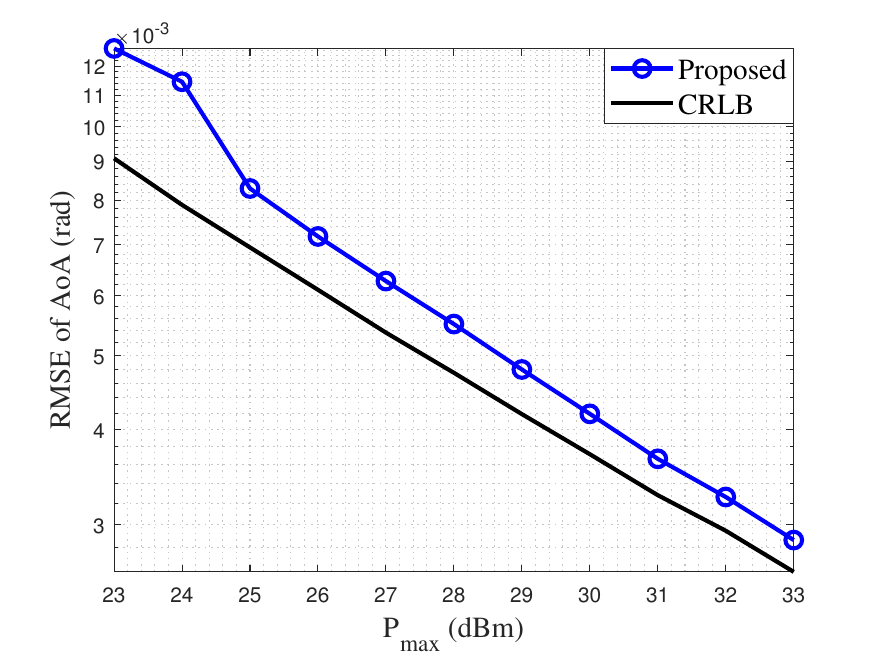}
		\footnotesize{(a) RMSE of AoA versus $P_{max}$. }
	\end{minipage}%
	\begin{minipage}[b]{0.33\linewidth}
		\centering
		\includegraphics[width=2.3in]{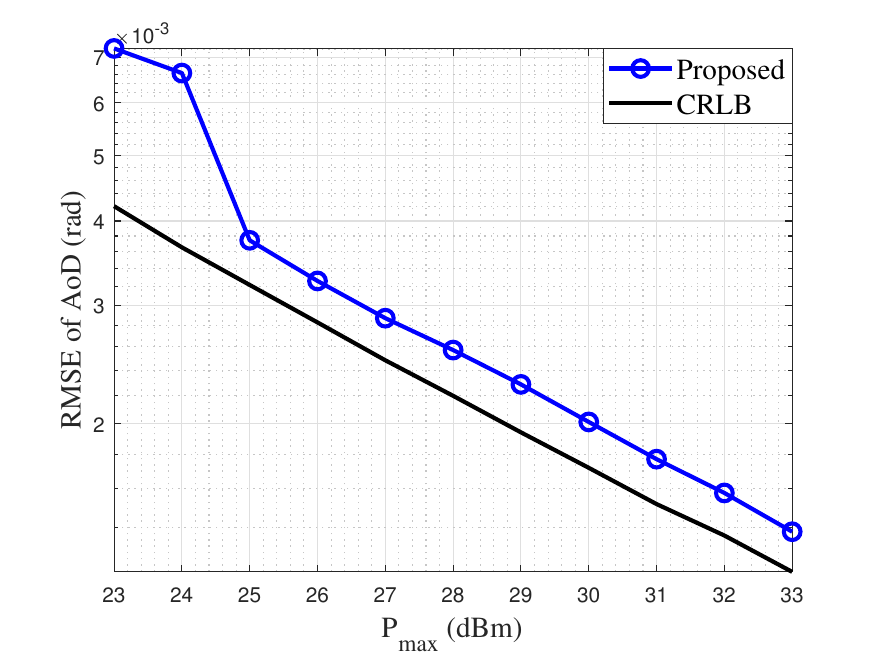}
		\footnotesize{(b) RMSE of AoD versus $P_{max}$.}
	\end{minipage}
	\begin{minipage}[b]{0.33\linewidth}
		\centering
		\includegraphics[width=2.3in]{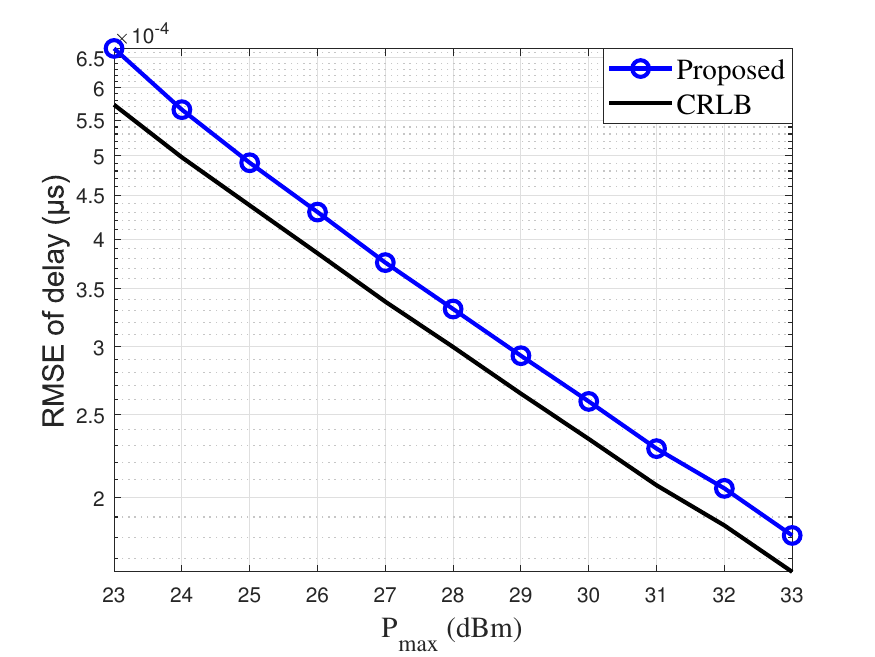}
		\footnotesize{(c) RMSE of delay versus $P_{max}$.}
	\end{minipage}
    \vspace{0.5em}
 \caption{RMSE performance of different parameters versus maximum transmit power with their corresponding CRLB. }
 \label{parameters_CRLB}
\end{figure*}


\section{Numerical Results}
In this section,  we provide numerical results to investigate the effectiveness of our proposed beamforming and target parameter estimation scheme for the \textcolor{black}{multi-hop RIS ISAC} system for home monitoring scenario. In the simulations, we assume that the Tx has $M_t=8$ antennas and is located at  $(0 \,\mathrm{m},15 \,\mathrm{m})$, the sensing Rx has $M_r=8$ antennas and located at $(0\,\mathrm{m},20\,\mathrm{m})$. \textcolor{black}{The number of RISs is set to be $I=2$.} The central point of RIS $1$ is located at $(-10 \,\mathrm{m},5 \,\mathrm{m})$ with $M_1=16$ reflecting elements, while the central point of RIS $2$ is located at $(0 \,\mathrm{m},0 \,\mathrm{m})$ with $M_2=16$ reflecting elements. For the communication users, we assume the number of users that are served by Tx directly is $K_0=3$, the number of users that are served by RIS $1$ is $K_1=2$ and the number of users that are served by RIS $2$ is $K_2=1$. Then the total number of communication users are $K=K_0+K_1+K_2=6$. \textcolor{black}{The communication users are randomly distributed in their corresponding areas and we assume that channel estimation has already been done for them in the sensing stage.
The number of sensing targets that need to be detected is $R=3$. The sensing area for the sensing Rx is set to be $20\, \mathrm{m} \times 20\,\mathrm{m}$, where the targets are randomly distributed in this area. The targets are assumed to be spatially separated and characterized by distinct parameters (i.e., AoA, AoD, and delay). For better illustration, the considered system geometry is shown in Fig.~\ref{simulation_setting}.}
\textcolor{black}{The carrier frequency $f_c$ is 28 GHz.}
The total bandwidth is $B= 20$ MHz \cite{bandwidth}, and the number of subcarriers is set to be $N=30$, which are equally divided among $K$ communication users, so that each user contains $J=5$ subcarriers.  The noise powers at the communication users and the sensing Rx are set to be $\sigma_c^2= -80$~dBm and $\sigma_s^2=- 80$~dBm \cite{noisepower}. The SNR threshold for each communication user is $\gamma_{thr}=20$~dB.
{Unless otherwise specified, the system parameters used in the simulations are summarized in Table~\ref{system_parameters}.}

\begin{table}[t]
\centering
\caption{\color{black}System Parameters}
\label{system_parameters}
\color{black}{\begin{tabular}{ll}
\toprule
\textbf{Parameter} & \textbf{Value} \\
\midrule
Carrier frequency $f_c$ & 28 GHz \\
Bandwidth $B$ & $20$ MHz \\
Number of subcarriers $N$ & $30$ \\
Number of Tx antennas $M_t$  & $8$ \\
Number of Rx antennas $M_r$  & $8$ \\
Number of RISs $I$ & $2$ \\
Number of RIS 1 reflecting elements $M_1$  & $16$ \\
Number of RIS 2 reflecting elements $M_2$  & $16$ \\
Number of users served by Tx directly $K_0$ & $3$\\
Number of users served by RIS 1 $K_1$ & $2$\\
Number of users served by RIS 2 $K_2$ & $1$\\
Total number of communication users $K$ & $6$\\
Number of sensing targets $R$ & $3$\\
Number of subcarriers at each communication users $J$ & 5\\ 
Noise power at the communication users $\sigma_c^2$ & $- 80$~dBm \\
Noise power at the sensing Rx $\sigma_s^2$ & $- 80$~dBm \\
SNR threshold for each communication user $\gamma_{thr}$ & $20$~dB\\
\bottomrule
\end{tabular}}
\end{table}

\subsection{Performance Evaluation of the Proposed Scheme}
In this subsection, we first provide the simulation results to verify the effectiveness of the proposed design. \textcolor{black}{To illustrate the fast convergence rate of the proposed scheme in terms of the required number of iterations, Fig.~\ref{iteration} illustrates the evolution of the NMSE performance of $ {\boldsymbol{\mathcal{G}}_{IR}}$ versus the number of iterations under different SNR conditions and different numbers of targets, which is defined as $\left\|{\boldsymbol{\mathcal{\hat G}}_{IR}-{\boldsymbol{\mathcal{G}}_{IR}}}\right\|_F^2/ \left\|{\boldsymbol{\mathcal{G}}_{IR}}\right\|_F^2$. It is observed that the NMSE decreases monotonically with the number of iterations and exhibits a rapid reduction during the first few iterations, followed by a gradual convergence to a stable value. As the transmit power increases, the effective sensing SNR is improved, allowing the proposed algorithm to converge to a lower NMSE value. Moreover, by employing a suitable initialization, a fast convergence behavior is observed. Although increasing the number of targets leads to a higher tensor rank and a higher probability of closely spaced parameters, which may slow down the convergence, the algorithm still converges rapidly within approximately 25 iterations. }

Since the final locations of the sensing targets are calculated based on AoA, AoD and delay, we then show the RMSE of the parameter estimation results. In Fig.~\ref{parameters_CRLB}, we depict the RMSE performance of AoA, AoD and delay of our proposed scheme compared with CRLB as a function of the maximum transmit power $P_{\max}$ at the transmitter. The SNR threshold at the communication user is set to $\gamma_{thr}=20$ dB. As the $P_{\max}$ increases, the RMSEs gradually decrease and approach their corresponding CRLBs, which exhibit an exponential decrease trend. This result further confirms the optimality of our proposed scheme. The optimality comes from the fact that ALS produces maximum likelihood estimation with i.i.d. Gaussian, which has been proved in \cite{low_rank} \cite{942635}.

\begin{figure*}
	\setlength{\abovecaptionskip}{-5pt}
	\setlength{\belowcaptionskip}{-10pt}
	\centering
	\begin{minipage}[b]{0.33\linewidth}
		\centering
		\includegraphics[width=2.3in]{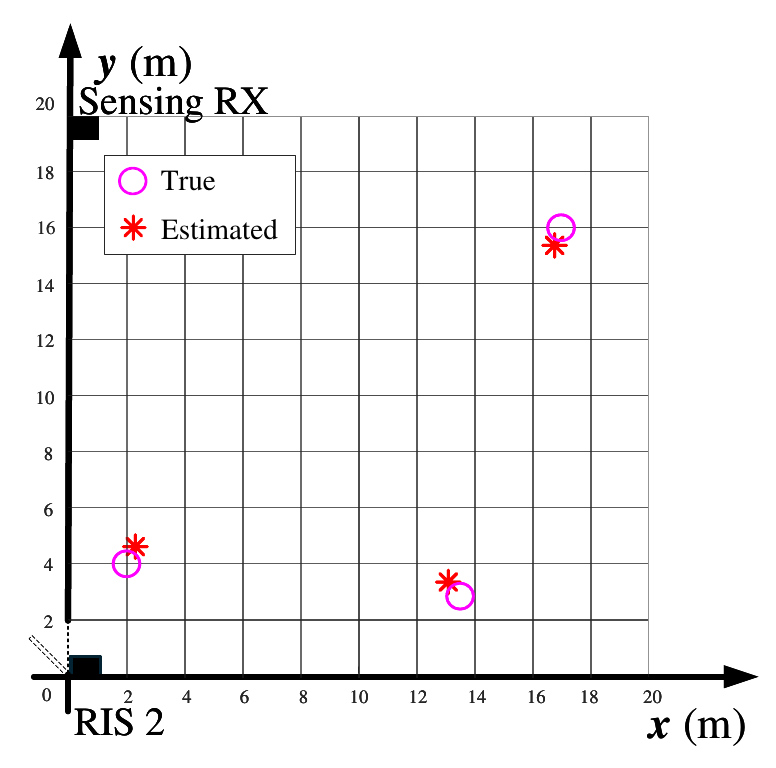}
		\footnotesize{(a) $P_\text{max}=23$ dBm}
		\label{fig3}
	\end{minipage}%
	\begin{minipage}[b]{0.33\linewidth}
		\centering
		\includegraphics[width=2.3in]{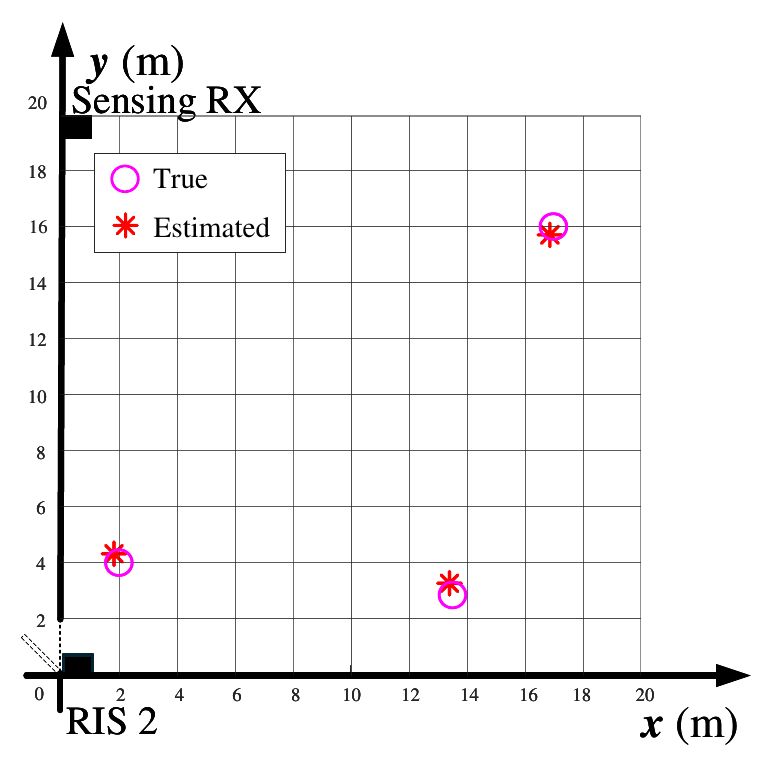}
		\footnotesize{(b) $P_\text{max}=28$ dBm}
		\label{fig4}
	\end{minipage}
	\begin{minipage}[b]{0.33\linewidth}
		\centering
		\includegraphics[width=2.3in]{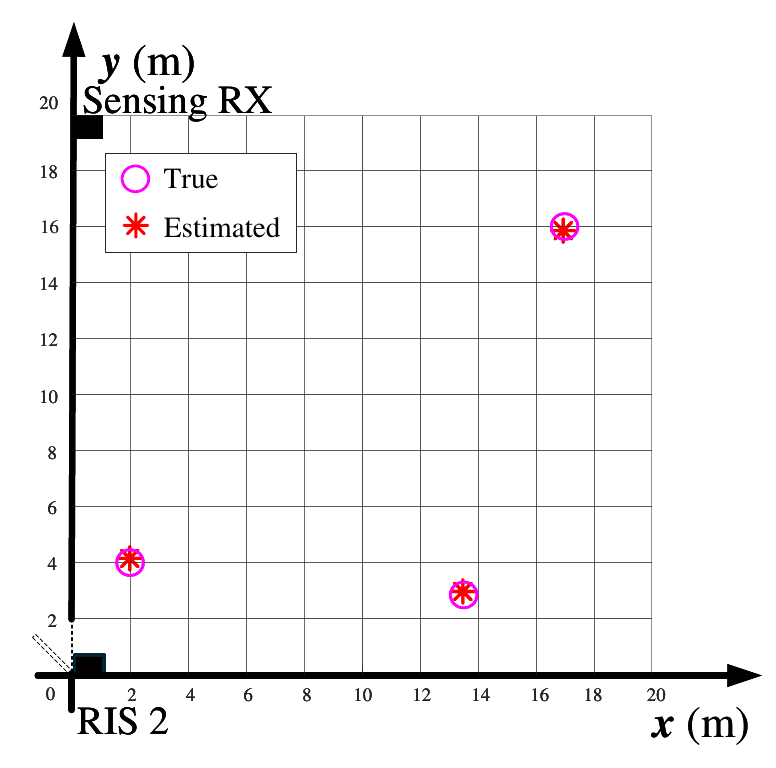}
		\footnotesize{(c) $P_\text{max}=33$ dBm}
		\label{fig5}
	\end{minipage}
    \vspace{0.5em}
 \caption{Comparison of the estimated position and the true position of the targets with different transmitted power $P_{max}$ at Tx.}
 \label{position}
\end{figure*}

{To intuitively illustrate the impact of maximum transmit power $P_\text{max}$ on localization performance, we present the localization results under different $P_\text{max}$ conditions in Fig.~\ref{position}. The numbers of RIS element are set to be $M_1=M_2=12$ and three targets located at $(2 \,\mathrm{m},4 \,\mathrm{m})$, $(17 \,\mathrm{m},16 \,\mathrm{m})$, and $(13.5 \,\mathrm{m},2.8 \,\mathrm{m})$ are taken as an example.} Since in this work we focus more on sensing the targets in the outdoor environment, we have omitted the indoor part on the left side of Fig.~\ref{simulation_setting}. When the energy allocated to communication user is held constant, the increase in $P_\text{max}$ results in improved target parameter estimation performance. As a result, the estimated positions of the targets converge increasingly towards their true positions.

\begin{figure}[h]
\centering
\includegraphics[width=\linewidth]{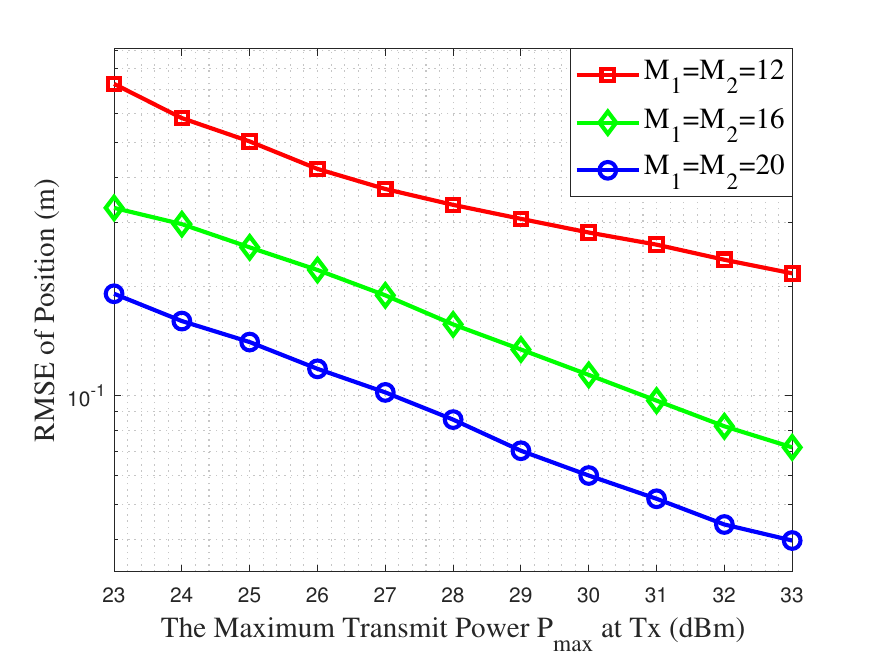}
\caption{RMSE performance of target positions versus maximim transmit power $P_{max}$ for different $M_1$ and  $M_2$ values. }
\label{RMSE_loc_SNR}
\end{figure}

In Fig.~\ref{RMSE_loc_SNR}, we further illustrate the RMSE performance of target localization with the estimated parameters versus the maximum transmit power $P_{\max}$ at the Tx side. The SNR threshold at the communication user is set to be $\gamma_{thr}=20$ dB, and the cases, $M_1=M_2=12$, $M_1=M_2=16$, and $M_1=M_2=20$ are used. The RMSE performance exhibits an exponential decreasing trend as the increasement of the maximum transmit power $P_\text{max}$ at the transmitter side. Additionally, a larger number of RIS reflecting elements allows for better energy focusing and provides higher passive reflecting gain, resulting in improved localization accuracy. 

\begin{figure}[t]
\centering
\includegraphics[width=\linewidth]{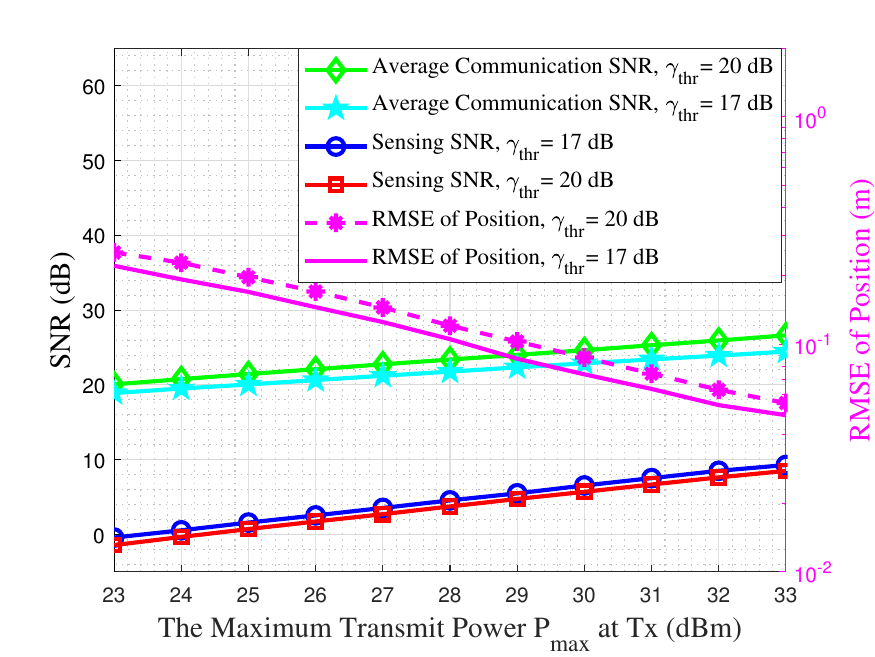}
\caption{\textcolor{black}{Received sensing SNR at the sensing Rx and the average received SNR for all communication users, as well as RMSE of target positions versus the transmit power $\mathrm{P_{max}}$.}}
\label{Comm_Eva}
\end{figure}

\begin{figure}[t]
 \vspace{-0.12in}
\centering
\includegraphics[width=\linewidth]{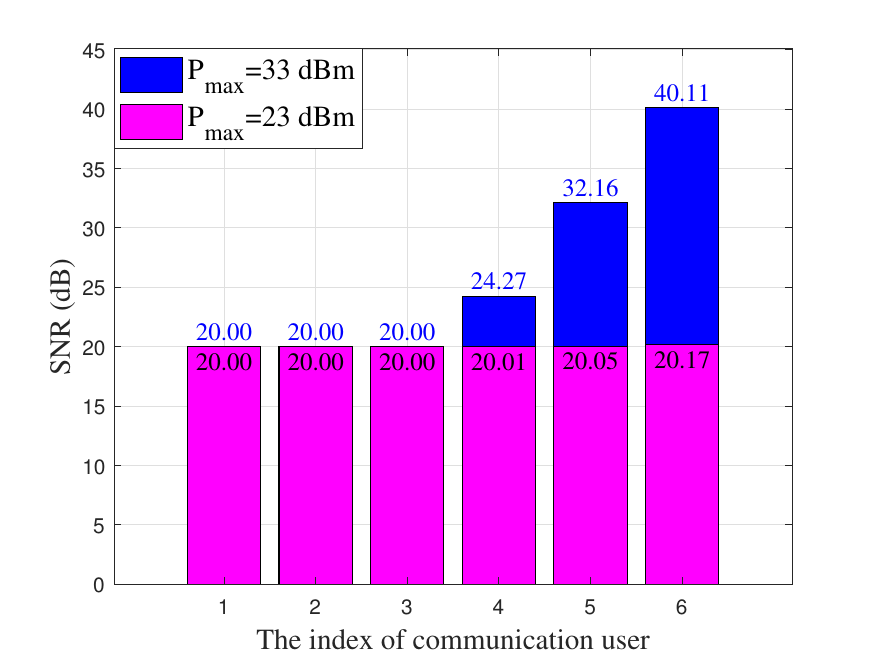}
\caption{\textcolor{black}{ The SNR of each communication user when the maximum transmit powers are $\mathrm{P_{max}}=23$~dBm and $\mathrm{P_{max}}=33$~dBm.}}
\label{Eachuser}
\end{figure}

\textcolor{black}{To further investigate the relationship between sensing and communication performances in this work,
in Fig.~\ref{Comm_Eva}, we show the SNRs for communication users and the sensing Rx and the corresponding sensing RMSE as the maximum transmit power $P_{\max}$ increases, under two different communication thresholds, i.e., $\gamma_{thr} = 20$~dB and $\gamma_{thr} = 17$~dB.
 As observed from the figure, increasing $P_{max}$ improves both the sensing and communication performance. Moreover, when a higher communication threshold ($\gamma_{thr} = 20$~dB) is imposed, the average SNR of the communication users increases, while the received sensing SNR decreases compared with the case of $\gamma_{thr} = 17$~dB. This result indicates a fundamental tradeoff between sensing and communication performance under limited transmit powers, and demonstrates how higher communication requirements can influence the performance of sensing.} 
 
\textcolor{black}{{Fig. \ref{Eachuser} shows the achieved SNRs of the six communication users with $P_{\max}=23$~dBm and $P_{\max}=33$~dBm. The communication threshold for each user is fixed as $\gamma_{{thr}}=20$~dB.} We can observe from the figure that some users have higher SNRs than the threshold because the constraint is $\gamma_{k,j} \ge \gamma_{thr}$, which only enforces a minimum requirement. Since the optimization maximizes the sensing power subject to the communication SNR constraints, the joint beamformings tend to steer energy toward the sensing target directions. As a result, communication users whose effective angles (AoD/AoA) are closer to the sensing directions can obtain significantly higher gains, leading to higher SNRs above the threshold.
}

These simulation results validate the effectiveness of our proposed approach. To demonstrate its superiority, we conduct the performance comparisons with other benchmark schemes in the next subsection.
\subsection{Performance Comparison with other Benchmark Schemes}
In this subsection, we compare our proposed tensor-decomposition based ISAC scheme with the following comparison benchmarks:
\begin{itemize}
    \item MUSIC-MF-based \cite{liu_fan_8999605} sensing scheme. Multiple Signal Classification (MUSIC) algorithm is a subspace-based super-resolution algorithm that utilizes the orthogonality between the signal subspace and the noise subspace. It is widely used in angle estimation. Matched-Filtering (MF) algorithm optimize signal detection and parameter estimation by correlating the received signal with a predefined signal template (filter). It is widely used in radar imaging and channel estimation. 
    
    \item BOMP-based \cite{9573305} sensing scheme. Block Orthogonal Matching Pursuit (BOMP) algorithm is designed for block-sparse signal recovery in the context of compressed sensing. BOMP leverages the special structure of block-sparse to improve recovery performance by selecting blocks of indices instead of individual indices. It is widely used in  channel estimation applications, particularly for mmWave massive MIMO wireless communication systems.
    \textcolor{black}{\item Random RIS phase shift design \cite{Random}. For the phase shift designs of RISs, we randomly generate $I_{ran}$ candidate phase-shift configurations and select the one that yields the minimum sensing RMSE while satisfying the communication constraints. } 
    \textcolor{black}{\item Separate transmit beamforming design \cite{Nonjoint}. The sensing and communication beams are designed independently. As a result, the total number of transmit beams equals the sum of the numbers of communication users and sensing targets. This benchmark represents a scenario where sensing and communication share the same hardware infrastructure, while their signal processing and beamforming are designed separately.}
\end{itemize}

\begin{table*}[t]
\centering
\caption{Dominant Computational Complexity Comparison} 
\label{tab:complexity}
{\color{black}
\renewcommand{\arraystretch}{1.4}
\begin{tabular}{|c|c|c|}
\hline
\textbf{Schemes} & \textbf{Beamforming complexity order} & \textbf{Sensing complexity order}\\ 
\hline
Proposed beamforming + proposed CPD-based sensing & $\mathcal{O}\big(KM_t^{3.5}+ M_1^{3.5}+M_2^{3.5}\big)$ & $\mathcal{O}\big(M_2M_rN(M_tQ+RI_{ite})\big)$ \\ 
\hline
Proposed beamforming + MUSIC-MF-based \cite{liu_fan_8999605} sensing & $\mathcal{O}\big(KM_t^{3.5}+ M_1^{3.5}+M_2^{3.5}\big)$ & $\mathcal{O}\big(M_r^2(R+G_\varphi)+M_r^3+M_2N G_\vartheta G_\tau\big)$ \\ 
\hline 
Proposed beamforming + BOMP-based \cite{9573305} sensing & $\mathcal{O}\big(KM_t^{3.5} + M_1^{3.5}+M_2^{3.5}\big)$ & $\mathcal{O}\big(M_2M_rNK_\varphi K_\vartheta K_\tau\big)$ \\ 
\hline
Random beamforming \cite{Random} + proposed CPD-based sensing & $\mathcal{O} \big(KM_t^{3.5} + I_{ran}M_2(M_1+2M_r) \big)$ & $\mathcal{O}\big(M_2M_rN(M_tQ+RI_{ite})\big)$\\
\hline
Separate beamforming \cite{Nonjoint} + proposed CPD-based sensing & $\mathcal{O} \big((K+R)M_t^{3.5} + M_1^{3.5}+M_2^{3.5}\big)$ & $\mathcal{O}\big(M_2M_rN(M_tQ+RI_{ite})\big)$\\
\hline
\end{tabular}
}
\end{table*}

The dominant computational complexity orders of the proposed scheme and all the benchmark schemes for our two-RIS setting are summarized in Table II, where $I_{ran}$ represents the number of random RIS configurations we generated in random RIS phase shift design scheme. For the complexity order of the sensing schemes, $G_\varphi$, $G_\vartheta$ and $G_\tau$ denote the numbers of search grids for AoA, AoD, and delay in the MUSIC-MF-based \cite{liu_fan_8999605} sensing scheme, respectively, while $K_\varphi$, $K_\vartheta$ and $K_\tau$ represent the codebook grid sizes for AoA, AoD, and delay in the BOMP-based \cite{9573305} sensing scheme. From Table II, we can observe that our proposed beamforming scheme has comparable complexity to the benchmarks. Meanwhile, compared with MUSIC-MF based sensing scheme, which involves a two-dimensional exhaustive search over the AoD and delay domain, and the BOMP-based sensing scheme that relies on a large codebook, our proposed CPD-based sensing scheme can have lower complexity and superior RMSE performance.

\begin{figure}
 \vspace{-0.12in}
\centering
\includegraphics[width=\linewidth]{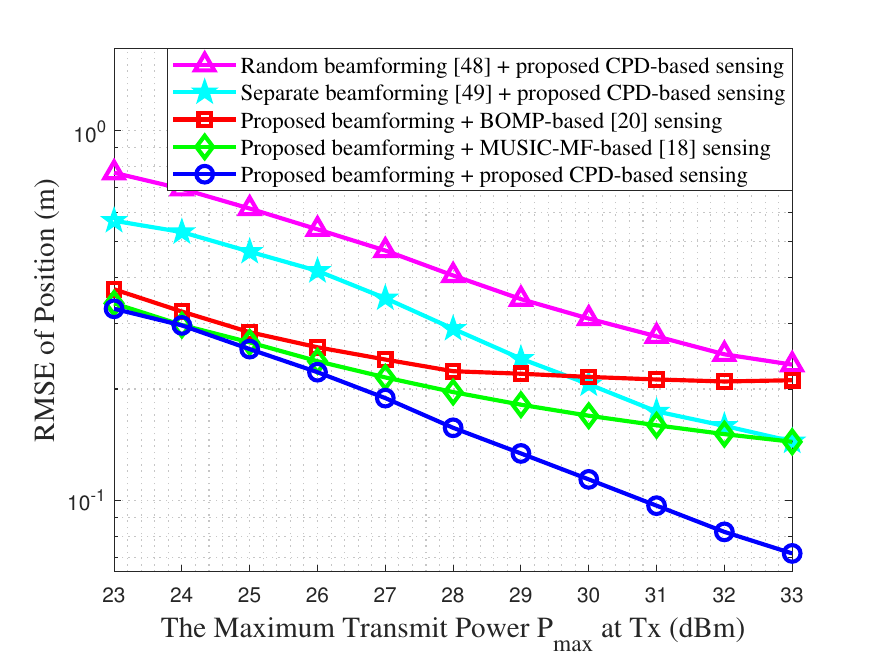}
\caption{\textcolor{black}{Comparison of RMSE performance versus maximum transmit power $P_{max}$ for different beamforming and sensing schemes.}}
\label{RMSE_power}
\end{figure}


\begin{figure}
\centering
\includegraphics[width=\linewidth]{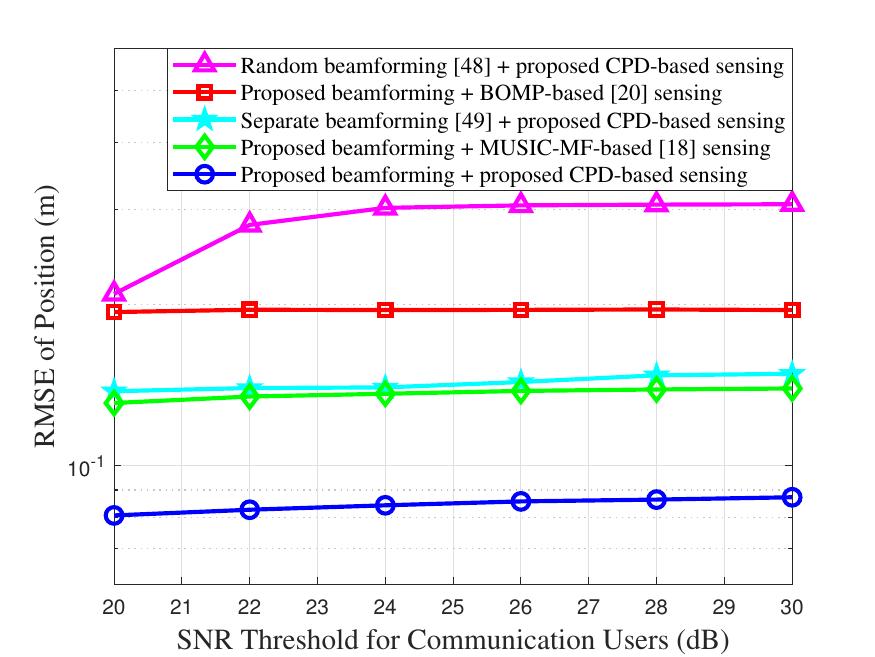}
\caption{\textcolor{black}{Comparison of RMSE performance versus communication threshold $\gamma_{\text{thr}}$ for different beamforming and sensing schemes.}}
\label{RMSE_threshold}
\end{figure}

\textcolor{black}{Fig.~\ref{RMSE_power} illustrates the RMSE performance of target position estimation versus the maximum transmit power $P_{\max}$ at the transmitter for different beamforming and sensing schemes. 
It can be observed that with increasing transmit power, the localization error decreases, but our proposed beamforming with proposed CPD-based sensing consistently achieves the lowest RMSE among all compared schemes. In contrast, the proposed beamforming schemes employing MUSIC-MF–based \cite{liu_fan_8999605} sensing and BOMP-based \cite{9573305} sensing exhibit inferior performance. Compared with the matrix-based MUSIC-MF scheme, the performance gain of our scheme comes from the ability to capture the intrinsic multi-dimensional structure of the sensing channel and to estimate the parameters independently, which helps to improve the performance. The BOMP-based sensing scheme exhibits poor RMSE performance in our scenario due to its inherent reliance on a predefined discretized grid, which limits its resolution and accuracy. In contrast, our proposed CPD-based sensing scheme operates in a fundamentally gridless manner \cite{low_rank}, thereby mitigating the quantization error and alleviating the constraints imposed by grid resolution. For the separate transmit beamforming scheme \cite{Nonjoint}, the sensing and communication beams are designed independently, which prevents the transmitter from exploiting the potential cooperation gain between sensing and communication. Moreover, allocating separate beams to sensing and communications reduces the available transmit power across all the beams. As a result, its RMSE performance is worse than the proposed joint sensing and communication schemes. For the random phase shift design scheme \cite{Random}, since the RIS reflection patterns are not optimized, the cascaded channels cannot be coherently aligned toward the sensing Rx. Consequently, the received signal strength for target localization is significantly reduced compared with the other schemes, leading to the worst RMSE performance in this simulation. }

Finally, we show how the RMSEs of our proposed method and all the benchmark schemes change with the SNR threshold $\gamma_{thr}$ on the communication user side in Fig.~\ref{RMSE_threshold}, where we set $P_{\max}=33$ dBm. We can observe that as the SNR threshold at the communication user $\gamma_{thr}$ increases, the localization performance of all methods deteriorates. This is due to the fact that, with a constant total transmission power $P_\text{max}$, an increase in the energy required for communication results in a reduction in the energy allocated to the sensing component, consequently lowering the sensing performance at the sensing Rx. Compared with other benchmark schemes, our proposed scheme can always achieve the lowest RMSE, and the sensing performance of our proposed scheme is insensitive to variations in the communication threshold $\gamma_{thr}$.
\vspace{-0.09in}
\subsection{\textcolor{black}{Performance Comparison under Imperfect Signal Cancellation Reflected by  Communication Users}}
\begin{figure}[t!]
\centering
\includegraphics[width=\linewidth]{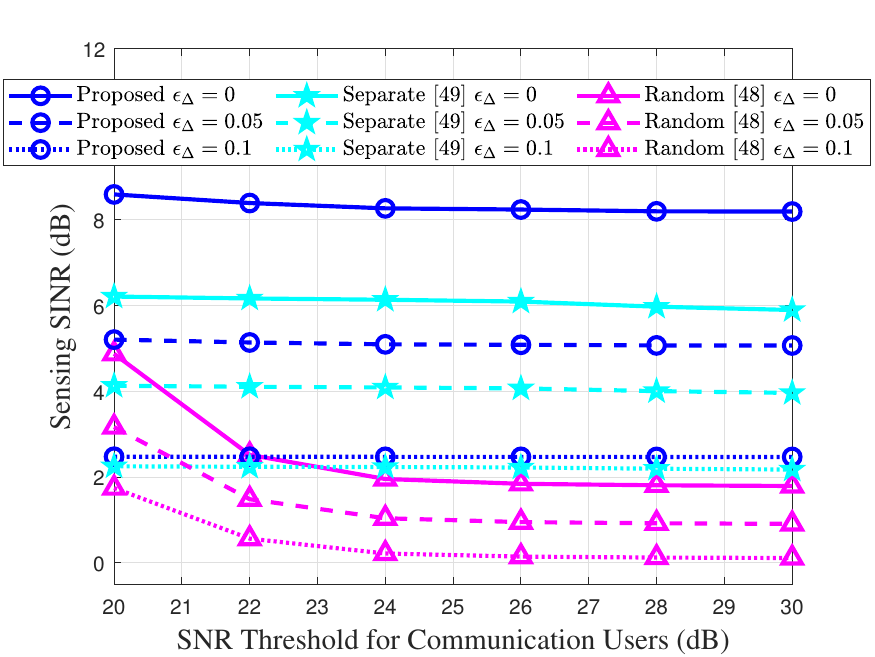}
\caption{\textcolor{black}{SINR of different beamforming schemes versus communication threshold $\gamma_{thr}$ for different residual cancellation factor $\epsilon_\Delta$.}}
\label{sensingSINR}
\end{figure}

\begin{figure}[t!]
\centering
\includegraphics[width=\linewidth]{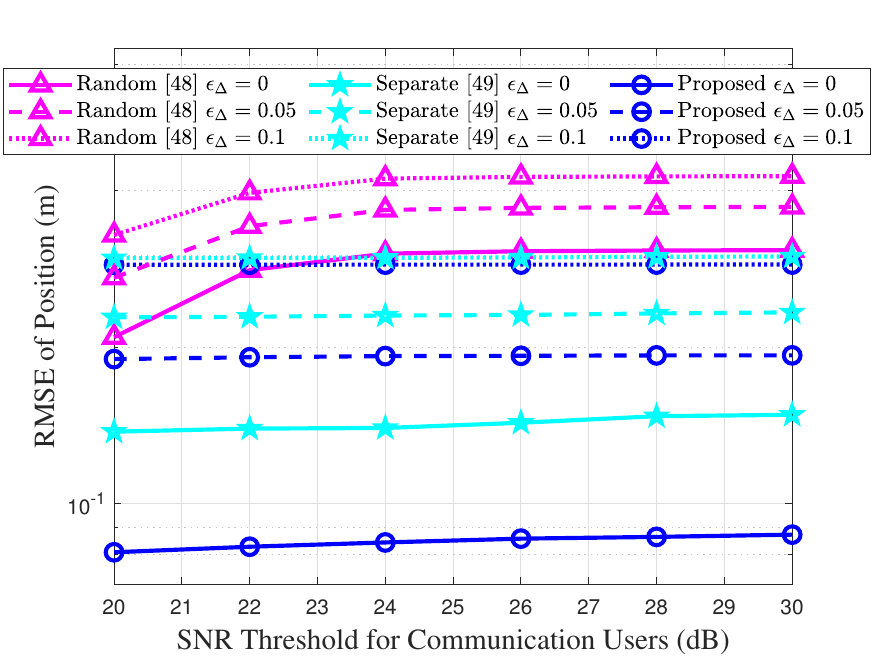}
\caption{\textcolor{black}{RMSE performance of different beamforming schemes versus communication threshold $\gamma_{thr}$ for different residual cancellation factor $\epsilon_\Delta$.}}
\label{sensingRMSEinter}
\end{figure}
\textcolor{black}{
In this subsection, we provide comparisons with benchmark beamforming schemes under imperfect signal cancellation of the {outdoor communication users}. Specifically, here we consider the case that in \eqref{have_re_in}, the cancellation is imperfect, i.e.,  $\mathbf{H}_{\text{COM},q,n}-\mathbf{\hat H}_{\text{COM},q,n}\neq \mathbf{0}$, and we model this reconstruction error as  $\Delta \mathbf{H}_{\text{COM},q,n} = \mathbf{H}_{\text{COM},q,n}-\mathbf{\hat H}_{\text{COM},q,n}$,
which follows a zero mean Gaussian distribution \cite{7518612, 8434253} of $\Delta \mathbf{H}_{\text{COM},q,n}\backsim \mathcal{CN}(\mathbf{0},\sigma_{\Delta}^2\boldsymbol{I})$.
Let the residual interference after the cancellation of the received signal reflected by active communication users as $\mathbf{E}_{q,n}=\left(\mathbf{H}_{\mathrm{COM},q,n}-\hat{\mathbf{H}}_{\mathrm{COM},q,n}\right)\mathbf{S}_{q,n} =\Delta\mathbf{H}_{\mathrm{COM},q,n}\mathbf{S}_{q,n}$, and $\mathbf{E}_{q,n}$ also follows a zero-mean complex Gaussian distribution, i.e., $ \mathbf{E}_{q,n}\sim\mathcal{CN}\left(\mathbf{0},\sigma_{\Delta}^{2}   \left\|\mathbf{S}_{q,n}\right\|^{2}\boldsymbol{I}
\right)$. We further define the effective noise term ${{\mathbf{\tilde N}}_{q,n}}$, which contains both the residual interference $\mathbf{E}_{q,n}$ and the additive noise ${\mathbf{N}}_{q,n}$, i.e, $ {{\mathbf{\tilde N}}_{q,n}}=\mathbf{E}_{q,n}+{\mathbf{N}}_{q,n}$, then ${{\mathbf{\tilde N}}_{q,n}}$ follows the following distribution ${{\mathbf{\tilde N}}_{q,n}}\backsim \mathcal{CN}\left(\mathbf{0},\left(\sigma^2_s+\sigma_{\Delta}^2\|{\mathbf{S}}_{q,n}\|^2\right)\boldsymbol{I}\right)$. We can observe that the variance of the effective noise ${{\mathbf{\tilde N}}_{q,n}}$ is directly affected by the transmitted communication signal power $\left\|\mathbf{S}_{q,n}\right\|^{2}$. In practical downlink transmission, this signal power is determined by the communication beamforming design and the SNR requirement of the communication user. Specifically, a higher SNR threshold generally requires the transmitter to allocate more beamforming power toward the communication link, which increases $\left\|\mathbf{S}_{q,n}\right\|^{2}$. To characterize the impact of imperfect reconstruction and cancellation in a normalized manner, we define a residual cancellation factor $\epsilon_\Delta$. Specifically, the variance of the reconstruction error is parameterized as $\sigma_{\Delta}^2=\epsilon_{\Delta}\sigma_{\text{COM}}^2$, where $\sigma_{\text{COM}}^2$ denotes the average power of channel for the active communication users and $\epsilon_{\Delta}$ denotes the residual cancellation factor. Smaller $\epsilon_{\Delta}$ means more accurate communication CSI and more effective cancellation.
Based on the above analysis, we compare the proposed beamforming scheme with the benchmark beamforming methods in terms of the sensing SINR (the ratio of the useful sensing signal power to the sum of the residual interference power and the noise power), and the resulting sensing RMSE performance. Here, we only focus on different beamforming schemes {with our proposed CPD-based sensing scheme}, since the residual interference is determined by the different beamforming methods and SNR requirements of the communication users. }

\textcolor{black}{
In Fig. \ref{sensingSINR}, we compare the sensing SINR achieved by different beamforming schemes, and Fig. \ref{sensingRMSEinter} shows the resulting sensing RMSE correspondingly. In the simulation, we consider three residual cancellation settings, i.e., $\epsilon_\Delta = 0$,
$\epsilon_\Delta = 0.05$ and $\epsilon_\Delta = 0.1$. As shown in Fig. \ref{sensingSINR}, the residual interference affects all considered beamforming schemes and reduces the sensing SINR compared with the ideal cancellation case ($\epsilon_\Delta = 0$). When $\epsilon_\Delta$ increases from $0.05$ to $0.1$, the sensing SINR decreases due to the stronger residual interference. Nevertheless, the proposed scheme consistently achieves the highest sensing SINR among all considered schemes under both residual cancellation settings. This demonstrates that, although residual interference degrades sensing performance, the proposed beamforming design can still provide the most favorable sensing SINR. This observation is further confirmed by the localization results in Fig. \ref{sensingRMSEinter}. As $\epsilon_\Delta$ increases, the RMSE generally increases, indicating that residual interference can degrade localization accuracy. However, the proposed scheme consistently achieves the lowest RMSE among all compared schemes for both $\epsilon_\Delta = 0.05$ and $\epsilon_\Delta = 0.1$. These results show that while residual interference has a noticeable impact on sensing and localization performance, the proposed scheme can still provide the best overall performance under practical residual cancellation errors.}

\section{Conclusion}
In this paper, we have investigated a novel \textcolor{black}{multi-hop RIS-assisted} ISAC system tailored for home security monitoring scenarios. By deploying multiple RISs, we enabled reliable communication between the Tx and associated users, while simultaneously improving the sensing Rx’s sensing capabilities for accurate target localization. To address the dual objectives, we formulated an optimization problem aimed at minimizing the RMSE of target detection under user communication constraints. {With maximized received sensing power at the sensing Rx, we unfolded the cascaded sensing channel and developed a low-rank CP decomposition-based approach to extract location parameters of the sensing target.} We also presented an analysis of the uniqueness conditions and derived CRLB for our estimation results. Extensive simulation results validated the effectiveness and superiority of the proposed scheme, demonstrating significant improvements in sensing performance while maintaining reliable communication service.

\bibliographystyle{IEEEtran}
\footnotesize
\bibliography{RIS_ref}

@ARTICLE{xiaoyan_9320618,
  author={Ma, Xiaoyan and Guo, Shuaishuai and Zhang, Haixia and Fang, Yuguang and Yuan, Dongfeng},
  journal={IEEE Trans. Wireless Commun.}, 
  title={Joint Beamforming and Reflecting Design in Reconfigurable Intelligent Surface-Aided Multi-User Communication Systems}, 
  year={2021},
  volume={20},
  number={5},
  pages={3269-3283},
  month={May},}

@ARTICLE{xiaoyan_9829192,
  author={Ma, Xiaoyan and Fang, Yuguang and Zhang, Haixia and Guo, Shuaishuai and Yuan, Dongfeng},
  journal={IEEE Trans. Wireless Commun.}, 
  title={Cooperative Beamforming Design for Multiple {RIS}-Assisted Communication Systems}, 
  year={2022},
  volume={21},
  number={12},
  pages={10949-10963},
  month={Dec.},}

@article{brinton2025key,
  title={Key focus areas and enabling technologies for {6G}},
  author={Brinton, Christopher G and Chiang, Mung and Kim, Kwang Taik and Love, David J and Beesley, Michael and Repeta, Morris and Roese, John and Beming, Per and Ekudden, Erik and Li, Clara and others},
  journal={IEEE Commun. Mag.},
  volume={63},
  number={3},
  pages={84--91},
  year={2025},
  publisher={IEEE}
}

@article{zhu2024enabling,
  title={Enabling intelligent connectivity: A survey of secure {ISAC} in {6G} networks},
  author={Zhu, Xiaoqiang and Liu, Jiqiang and Lu, Lingyun and Zhang, Tao and Qiu, Tie and Wang, Chunpeng and Liu, Yuan},
  journal={IEEE Commun. Surv. Tuts.},
  year={2025},
   volume={27},
  number={2},
  pages={748-781},
  month={Apr.},
}

@article{cao2024sensing,
  title={Sensing for Secure Communication in {ISAC}: Protocol Design and Beamforming Optimization},
  author={Cao, Yang and Duan, Lingjie and Zhang, Rui},
  journal={IEEE Trans. Wireless Commun.},
  year={2025},
  volume={24},
  number={2},
  pages={1207-1220},
  month={Feb.},
}

@article{su2023sensing,
  title={Sensing-assisted eavesdropper estimation: An {ISAC} breakthrough in physical layer security},
  author={Su, Nanchi and Liu, Fan and Masouros, Christos},
  journal={IEEE Trans. Wireless Commun.},
  volume={23},
  number={4},
  pages={3162--3174},
  year={2023},
  publisher={IEEE}
}

@ARTICLE{GAN_10543050,
  author={Gan, Xu and Huang, Chongwen and Yang, Zhaohui and Zhong, Caijun and Chen, Xiaoming and Zhang, Zhaoyang and Guo, Qinghua and Yuen, Chau and Debbah, Mérouane},
  journal={IEEE J. Sel. Topics Signal Process.}, 
  title={Bayesian Learning for Double-{RIS} Aided {ISAC} Systems With Superimposed Pilots and Data}, 
  year={2024},
  volume={18},
  number={5},
  pages={766-781},
  month={Jul.},}

@article{Kruskal,
  author = {J. B. Kruskal},
  title = {Three-way arrays: Rank and uniqueness or trilinear decompositions, with applications to arithmetic complexity and statistics},
  journal = {Linear Algebra Appl.},
  volume = {18},
  number = {2},
  pages = {95--138},
  year = {1977}
}

@article{van_cpd,
  author = {M. Sørensen and L. De Lathauwer},
  title = {Blind signal separation via tensor decomposition with Vandermonde factor: Canonical polyadic decomposition},
  journal = {IEEE Trans. Signal Process.},
  volume = {61},
  number = {22},
  pages = {5507--5519},
  month = {Aug.},
  year = {2013}
}

@ARTICLE{942635,
  author={Xiangqian Liu and Sidiropoulos, N.D.},
  journal={IEEE Trans. Signal Process.}, 
  title={Cramer-Rao lower bounds for low-rank decomposition of multidimensional arrays}, 
  year={2001},
  volume={49},
  number={9},
  pages={2074-2086},
  month={Sep.},}

@ARTICLE{9573305,
  author={Mohebbi, Ali and Abdzadeh-Ziabari, Hamed and Zhu, Wei-Ping and Ahmad, M. Omair},
  journal={IEEE Trans. Veh. Technol.}, 
  title={Doubly Selective Channel Estimation Algorithms for Millimeter Wave Hybrid {MIMO} Systems}, 
  year={2021},
  volume={70},
  number={12},
  pages={12821-12835},
  month={Dec.},}

@ARTICLE{liu_fan_8999605,
  author={Liu, Fan and Masouros, Christos and Petropulu, Athina P. and Griffiths, Hugh and Hanzo, Lajos},
  journal={IEEE Trans. Commun.}, 
  title={Joint Radar and Communication Design: Applications, State-of-the-Art, and the Road Ahead}, 
  volume={68},
  number={6},
  pages={3834-3862},
  month={Jun.},
 year={2020},}

@ARTICLE{80211_ad_8114253,
  author={Kumari, Preeti and Choi, Junil and González-Prelcic, Nuria and Heath, Robert W.},
  journal={IEEE Trans. Veh. Technol.}, 
  title={{IEEE} 802.11ad-Based Radar: An Approach to Joint Vehicular Communication-Radar System}, 
  year={2018},
  volume={67},
  number={4},
  pages={3012-3027},
  month={Apr.},}

@ARTICLE{superresolution_7833233,
  author={Zheng, Le and Wang, Xiaodong},
  journal={IEEE Trans. Signal Process.}, 
  title={Super-Resolution Delay-{D}oppler Estimation for {OFDM} Passive Radar}, 
  year={2017},
  volume={65},
  number={9},
  pages={2197-2210},
  month={May},}

@ARTICLE{THz_9967989,
  author={Wu, Yongzhi and Lemic, Filip and Han, Chong and Chen, Zhi},
  journal={IEEE Trans. Commun.}, 
  title={Sensing Integrated {DFT}-Spread {OFDM} Waveform and Deep Learning-Powered Receiver Design for Terahertz Integrated Sensing and Communication Systems}, 
  year={2023},
  volume={71},
  number={1},
  pages={595-610},
  month={Jan.},}

@article{low_rank,
  author = {Z. Zhou and J. Fang and L. Yang and H. Li and Z. Chen and R. S. Blum},
  title = {Low-Rank Tensor Decomposition-Aided Channel Estimation for Millimeter Wave {MIMO-OFDM} Systems},
  journal = {IEEE J. Sel. Areas Commun.},
  volume = {35},
  number = {7},
  pages = {1524--1538},
  month = {Jul.},
  year = {2017}
}

@article{Nguyen2022,
  author={Nguyen, Dinh C. and Ding, Ming and Pathirana, Pubudu N. and Seneviratne, Aruna and Li, Jun and Niyato, Dusit and Dobre, Octavia and Poor, H. Vincent},
  title = {6{G} Internet of Things: A comprehensive survey},
  journal = {IEEE Internet Things J.},
  volume = {9},
  number = {1},
  pages = {359--383},
  month = {Jan.},
  year = {2022}
}

@article{Z_Zhang2019,
  author = {Z. Zhang and Y. Xiao and Z. Ma and M. Xiao and Z. Ding and X. Lei and G. K. Karagiannidis and P. Fan},
  title = {6{G} wireless networks: vision, requirements, architecture, and key technologies},
  journal = {IEEE Veh. Technol. Mag.},
  volume = {14},
  number = {3},
  pages = {28--41},
  month = {Sep.},
  year = {2019}
}

@article{Q_Qi2022,
  author = {Q. Qi and X. Chen and A. Khalili and C. Zhong and Z. Zhang and D. W. K. Ng},
  title = {Integrating sensing, computing, and communication in 6{G} wireless networks: Design and optimization},
  journal = {IEEE Trans. Commun.},
  volume = {70},
  number = {9},
  pages = {6212--6227},
  month = {Sep.},
  year = {2022}
}

@ARTICLE{10319318,
  author={Yu, Zhiyuan and Ren, Hong and Pan, Cunhua and Zhou, Gui and Wang, Boshi and Dong, Mianxiong and Wang, Jiangzhou},
  journal={IEEE Trans. Wireless Commun.}, 
  title={Active {RIS}-Aided {ISAC} Systems: Beamforming Design and Performance Analysis}, 
  year={2024},
  volume={72},
  number={3},
  pages={1578-1595},
  month={Mar.},}

@article{ruoyu_ISAC,
  author = {Zhang, Ruoyu and Cheng, Lei and Wang, Shuai and Lou, Yi and Gao, Yulong and Wu, Wen and Ng, Derrick Wing Kwan},
  title = {Integrated sensing and communication with massive {MIMO}: A unified tensor approach for channel and target parameter estimation},
  journal = {IEEE Trans. Wireless Commun.},
  volume = {23},
  number = {8},
  pages = {8571--8587},
  month={Aug.},
  year = {2024}
}

@ARTICLE{9540344,
  author={Zhang, J. Andrew and Liu, Fan and Masouros, Christos and Heath, Robert W. and Feng, Zhiyong and Zheng, Le and Petropulu, Athina},
  journal={IEEE J. Sel. Topics Signal Process.}, 
  title={An Overview of Signal Processing Techniques for Joint Communication and Radar Sensing}, 
  year={2021},
  volume={15},
  number={6},
  pages={1295-1315},
  month={Nov.},}

@ARTICLE{Rihan_10194901,
  author={Rihan, Mohamed and Zappone, Alessio and Buzzi, Stefano},
  journal={IEEE Trans. Commun.}, 
  title={Robust {RIS}-Assisted {MIMO} Communication-Radar Coexistence: Joint Beamforming and Waveform Design}, 
  year={2023},
  volume={71},
  number={11},
  pages={6647-6661},
  month={Nov.},}

@ARTICLE{Novel_10634583,
  author={Xiao, Zichao and Liu, Rang and Li, Ming and Liu, Qian and Swindlehurst, A. Lee},
  journal={IEEE Trans. Signal Process.}, 
  title={A Novel Joint Angle-Range-Velocity Estimation Method for {MIMO-OFDM} {ISAC} Systems}, 
  year={2024},
  volume={72},
  number={},
  pages={3805-3818},
  month={Aug.},}

@ARTICLE{OFDM_5776640,
  author={Sturm, Christian and Wiesbeck, Werner},
  journal={Proceedings of the IEEE}, 
  title={Waveform Design and Signal Processing Aspects for Fusion of Wireless Communications and Radar Sensing}, 
  year={2011},
  volume={99},
  number={7},
  pages={1236-1259},
  month={Jul.},}

@inproceedings{yirui_10200722,
  author = {Y. Luo and Y. Guan and E. Gunawan},
  title = {Uplink sensing with unknown transmitter position in clutter environment via tensor decomposition},
  booktitle = {Proc. IEEE 97th Veh. Technol. Conf. (VTC-Spring)},
  month = {Jun.},
  year = {2023},
  pages = {1--5}
}

@ARTICLE{yongjun_9166743,
  author={Liu, Yongjun and Liao, Guisheng and Chen, Yufeng and Xu, Jingwei and Yin, Yingzeng},
  journal={IEEE Trans. Veh. Technol.}, 
  title={Super-Resolution Range and Velocity Estimations With {OFDM} Integrated Radar and Communications Waveform}, 
  year={2020},
  volume={69},
  number={10},
  pages={11659-11672},
  month={Oct.},}

@ARTICLE{RIS1,
  author={Liu, Rang and Li, Ming and Liu, Qian and Lee Swindlehurst, A.},
  journal={IEEE Trans. Wireless Commun.}, 
  title={{SNR/CRB}-Constrained Joint Beamforming and Reflection Designs for {RIS-ISAC} Systems}, 
  year={2024},
  volume={23},
  number={7},
  pages={7456-7470},
  month={Jul.}, }

@ARTICLE{RIS2,
  author={Luo, Honghao and Liu, Rang and Li, Ming and Liu, Yang and Liu, Qian},
  journal={IEEE Trans. Veh. Technol.}, 
  title={Joint Beamforming Design for {RIS}-Assisted Integrated Sensing and Communication Systems}, 
  year={2022},
  volume={71},
  number={12},
  pages={13393-13397},
  month={Dec.}}

@ARTICLE{RIS3,
  author={Zhang, Jifa and Liu, Mingqian and Tang, Jie and Zhao, Nan and Niyato, Dusit and Wang, Xianbin},
  journal={IEEE Trans. Cognit. Commun. Networking}, 
  title={Joint Design for {RIS}-Aided {ISAC} via Deep Unfolding Learning}, 
  year={2025},
  volume={11},
  number={1},
  pages={349-361},
  month={Feb.}}

@ARTICLE{MYIOTJ10963873,
  author={Luo, Yirui and Guan, Yong Liang and Ge, Yao and González G, David and Yuen, Chau},
  journal={Internet Things J.}, 
  title={A Novel Angle-Delay-Doppler Estimation Scheme for {AFDM-ISAC} System in Mixed Near-Field and Far-Field Scenarios}, 
  year={2025},
  volume={12},
  number={13},
  pages={22669-22682},
  month={Jul.},}

@ARTICLE{Power1,
  author={Zhu, Qi and Li, Ming and Liu, Rang and Liu, Qian},
  journal={IEEE Trans. Wireless Commun.}, 
  title={Cramér-Rao Bound Optimization for Active {RIS}-Empowered {ISAC} Systems}, 
  year={2024},
  volume={23},
  number={9},
  pages={11723-11736},
  month={Sept.}}

@ARTICLE{Power2,
  author={He, Zhenyao and Shen, Hong and Xu, Wei and Eldar, Yonina C. and You, Xiaohu},
  journal={IEEE Trans. Signal Process.}, 
  title={{MSE}-Based Training and Transmission Optimization for {MIMO ISAC} Systems}, 
  year={2024},
  volume={72},
  number={},
  pages={3104-3121},
 month={Jun.}}

@ARTICLE{SDR,
  author={Luo, Zhi-{Q}uan and Ma, Wing-{K}in and So, Anthony Man-{C}ho and Ye, Yinyu and Zhang, Shuzhong},
  journal={IEEE Signal Process. Mag.}, 
  title={Semidefinite Relaxation of Quadratic Optimization Problems}, 
  year={2010},
  volume={27},
  number={3},
  pages={20-34},
  month={May}}

@ARTICLE{Guassian_Random,
  author={Wu, Qingqing and Zhang, Rui},
  journal={IEEE Trans. Wireless Commun.}, 
  title={Intelligent Reflecting Surface Enhanced Wireless Network via Joint Active and Passive Beamforming}, 
  year={2019},
  volume={18},
  number={11},
  pages={5394-5409},
  month={Nov.}}

@article{gap,
  title={On approximating complex quadratic optimization problems via semidefinite programming relaxations},
  author={So, Anthony Man-Cho and Zhang, Jiawei and Ye, Yinyu},
  journal={Mathematical Programming},
  volume={110},
  number={1},
  pages={93--110},
  year={2007},
  publisher={Springer},
month={Dec.}
}

@ARTICLE{Lee_7458188,
  author={Lee, Junho and Gil, Gye-Tae and Lee, Yong H.},
  journal={IEEE Trans. Commun.}, 
  title={Channel Estimation via Orthogonal Matching Pursuit for Hybrid {MIMO} Systems in Millimeter Wave Communications}, 
  year={2016},
  volume={64},
  number={6},
  pages={2370-2386},
  month={Jun.}}

@ARTICLE{bandwidth,
  author={Li, Chenglong and De Bast, Sibren and Tanghe, Emmeric and Pollin, Sofie and Joseph, Wout},
  journal={IEEE Sensors Journal}, 
  title={Toward Fine-Grained Indoor Localization Based on Massive {MIMO-OFDM} System: Experiment and Analysis}, 
  year={2022},
  volume={22},
  number={6},
  pages={5318-5328},
   month={Mar.}}

@article{tensor_Kolda,
  author    = {T. G. Kolda and B. W. Bader},
  title     = {Tensor decompositions and applications},
  journal   = {SIAM Review},
  volume    = {51},
  number    = {3},
  pages     = {455--500},
  year      = {2009},
  month     = {Sep.},
}

@ARTICLE{li10438390,
  author={Li, Yueheng and de Oliveira, Lucas Giroto and Diewald, Axel and Long, Xueyun and Bekker, Elizabeth and Brunner, David and Wan, Xiang and Cui, Tie Jun and Zwick, Thomas and Nuss, Benjamin},
  journal={IEEE Trans. Wireless Commun.}, 
  title={User Detection in {RIS}-Based mm{W}ave {JCAS}: Concept and Demonstration}, 
  year={2024},
  volume={23},
  number={8},
  pages={9596-9612},
  month={Aug.},}

@ARTICLE{chenjie10053657,
  author={Chen, Jie and Liang, Ying-Chang and Cheng, Hei Victor and Yu, Wei},
  journal={IEEE Trans. Wireless Commun.}, 
  title={Channel Estimation for Reconfigurable Intelligent Surface Aided Multi-User mmWave {MIMO} Systems}, 
  year={2023},
  volume={22},
  number={10},
  pages={6853-6869},
  month={Oct.},}

@ARTICLE{FarF0,
  author={Chai, Zhi and Xu, Jiajie and Coon, Justin P. and Alouini, Mohamed-Slim},
  journal={{\color{black}IEEE Trans. Veh. Technol.}}, 
  title={{\color{black}{RIS}-assisted millimeter wave communications for indoor scenarios: Modeling and coverage analysis}}, 
  year={2025},
  volume={},
  number={},
  pages={1-16}
}

@ARTICLE{FarF1,
  author={Ju, Shihao and Xing, Yunchou and Kanhere, Ojas and Rappaport, Theodore S.},
  journal={{\color{black}IEEE J. Sel. Areas Commun.}}, 
  title={{\color{black}Millimeter Wave and Sub-Terahertz Spatial Statistical Channel Model for an Indoor Office Building}}, 
  year={2021},
  volume={39},
  number={6},
  pages={1561-1575},
  month={Jun.}
}

@ARTICLE{LoSR,
  author={Maccartney, George R. and Rappaport, Theodore S. and Sun, Shu and Deng, Sijia},
  journal={{\color{black}IEEE Access}}, 
  title={{\color{black}Indoor office wideband millimeter-wave propagation measurements and channel models at 28 and 73 {GHz} for ultra-dense {5G} wireless networks}}, 
  year={2015},
  volume={3},
  number={},
  pages={2388-2424}}

@ARTICLE{9815098,
  author={Zhang, Ruoyu and Cheng, Lei and Wang, Shuai and Lou, Yi and Wu, Wen and Ng, Derrick Wing Kwan},
  journal={{\color{black}IEEE Trans. Commun.}}, 
  title={{\color{black}Tensor Decomposition-Based Channel Estimation for Hybrid mmWave Massive {MIMO} in High-Mobility Scenarios}}, 
  year={2022},
  volume={70},
  number={9},
  pages={6325-6340},
  month={Sep.},}

@ARTICLE{Nonjoint,
  author={Zhu, Zhengyu and Li, Zheng and Chu, Zheng and Guan, Yingying and Wu, Qingqing and Xiao, Pei and Renzo, Marco Di and Lee, Inkyu},
  journal={\textcolor{black}{IEEE Internet Things J.}}, 
  title={\textcolor{black}{Intelligent reflecting surface assisted {mmWave} integrated sensing and communication systems}}, 
  year={2024},
  volume={11},
  number={18},
  pages={29427-29437},
  ISSN={2327-4662},
  month={Sep.},}

@ARTICLE{Random,
  author={Yan, Han and Chen, Hua and Liu, Wei and Yang, Songjie and Wang, Gang and Yuen, Chau},
  journal={{\color{black}IEEE Trans. Green Commun. Netw.}}, 
  title={{\color{black}{RIS}-Enabled Joint Near-Field 3{D} Localization and Synchronization in SISO Multipath Environments}}, 
  year={2025},
  volume={9},
  number={1},
  pages={367-379},
  month={Mar.}}

@ARTICLE{noisepower,
  author={Hao, Wanming and Qu, Yongchao and Zhou, Shuang and Li, Xingwang and Zhu, Zhengyu and Yang, Liang},
  journal={\textcolor{black}{IEEE Trans. Wireless Commun.}}, 
  title={{\textcolor{black}{Secure energy efficiency optimization for sub-connected active RIS-assisted mmWave ISAC system}}}, 
  year={2026},
  volume={25},
  number={},
  pages={3960-3977}}

@ARTICLE{11506510,
  author={{Chen, Yaxuan and Zhang, Guangchi and Cui, Miao and Fu, Hao and Wu, Qingqing and Zhang, Rui}},
  journal={{IEEE Trans. Wireless Commun.}}, 
  title={{Sensing-Assisted Secure Communication in MA-Aided ISAC: CRB Analysis and Robust Design}}, 
  year={2026},
  volume={25},
  number={},
  pages={16400-16416},
  month={},}

@ARTICLE{7518612,
  author={Xiong, Xin and Wang, Xiaodong and Riihonen, Taneli and You, Xiaohu},
  journal={\textcolor{black}{IEEE Trans. Wireless Commun.}}, 
  title={\textcolor{black}{Channel estimation for full-duplex relay systems with large-scale antenna arrays}}, 
  year={2016},
  volume={15},
  number={10},
  pages={6925-6938},
  month={Oct.},}

@ARTICLE{8434253,
  author={Li, Xiaofeng and Tepedelenlioğlu, Cihan and Şenol, Habib},
  journal={{\color{black}IEEE Trans. Commun.}}, 
  title={\textcolor{black}{Optimal Training for Residual Self-Interference for Full-Duplex One-Way Relays}}, 
  year={2018},
  volume={66},
  number={12},
  pages={5976-5989},
  month={Dec.},}

\end{document}